\documentclass[a4paper,superscriptaddress,twocolumn,prb,showkeys,preprintnumbers,floatfix,showpacs]{revtex4-1}

\usepackage{epsfig}
\usepackage{amsmath}
\usepackage{amssymb}
\usepackage{blkarray, multirow, graphicx, diagbox, color, xcolor, colortbl}
\usepackage{ulem} 
\newcommand{\hopping}[0]{t_{\mathrm{h}}} 

\usepackage{braket}
\usepackage{tabularx}
\usepackage{booktabs} 
\newcolumntype{C}{>{$}c<{$}}
\AtBeginDocument{
	\heavyrulewidth=.08em
	\lightrulewidth=.05em
	\cmidrulewidth=.03em
	\belowrulesep=.65ex
	\belowbottomsep=0pt
	\aboverulesep=.4ex
	\abovetopsep=0pt
	\cmidrulesep=\doublerulesep
	\cmidrulekern=.5em
	\defaultaddspace=.5em
}

\usepackage{ifthen}
\usepackage{ulem}
\usepackage{tikz}
\usepackage{calc}
\usetikzlibrary{external}
\usepackage{pgffor}
\usepackage{pgfplots}
\pgfplotsset{compat=1.8}
\usepackage{pgfplotstable}
\usepgfplotslibrary{groupplots}

\definecolor{colorA}{rgb} {0.58,0,0.8275}
\definecolor{colorB}{rgb} {0.11,0.663,0.51}
\definecolor{colorC}{rgb} {0.3373,0.7059,0.9137}
\definecolor{colorD}{rgb} {0.902,0.6235,0}
\definecolor{colorE}{rgb} {0.9451,0.902,0.3255}

\usetikzlibrary
{
	calc,
	patterns,
	positioning,
	petri,
	arrows,
	decorations.markings,
	backgrounds,
	fit,
	graphs,
	shapes.geometric,
	decorations.pathmorphing,
	shapes.misc,
	shapes,
	tikzmark
}

\newcommand{\bspWidth}{0.6}
\newcommand{\bspArrowSize}{0.4}
\newcommand{\Zeeman}{0.75}
\newcommand{\yshift}[1]
{
	ifthenelse(mod(#1,2)<1,0,1)*\Zeeman
}

\newcommand{\bspNone}[1]
{
	\draw (#1*\bspWidth*1.5,{\yshift{#1}}) -- coordinate[midway] (m#1) (#1*\bspWidth*1.5+\bspWidth,{\yshift{#1}});
}
\newcommand{\bspGhost}[1]
{
	\draw[white] (#1*\bspWidth*1.5,{\yshift{#1}}) -- coordinate[midway] (m#1) (#1*\bspWidth*1.5+\bspWidth,{\yshift{#1}});
}
\newcommand{\bspUp}[1]
{
	\draw (#1*\bspWidth*1.5,{\yshift{#1}}) --  (#1*\bspWidth*1.5+\bspWidth,{\yshift{#1}});
	\draw[->, thick] (#1*\bspWidth*1.5+0.5*\bspWidth,{\yshift{#1}-\bspArrowSize*0.5}) --  (#1*\bspWidth*1.5+0.5*\bspWidth,{\yshift{#1}+\bspArrowSize*0.5}) coordinate (m#1);
}
\newcommand{\bspDown}[1]
{
	\draw (#1*\bspWidth*1.5,{\yshift{#1}}) -- (#1*\bspWidth*1.5+\bspWidth,{\yshift{#1}});
	\draw[->, thick] (#1*\bspWidth*1.5+0.5*\bspWidth,{\yshift{#1}+\bspArrowSize*0.5}) coordinate (m#1) -- (#1*\bspWidth*1.5+0.5*\bspWidth,{\yshift{#1}-\bspArrowSize*0.5});
}
\newcommand{\bspUpDown}[1]
{
	\draw (#1*\bspWidth*1.5,{\yshift{#1}}) -- (#1*\bspWidth*1.5+\bspWidth,{\yshift{#1}});
	\draw[->, thick] (#1*\bspWidth*1.5+0.3*\bspWidth,{\yshift{#1}-\bspArrowSize*0.5}) -- (#1*\bspWidth*1.5+0.3*\bspWidth,{\yshift{#1}+\bspArrowSize*0.5}) coordinate (m#1a); 
	\draw[->, thick] (#1*\bspWidth*1.5+0.7*\bspWidth,{\yshift{#1}+\bspArrowSize*0.5}) coordinate (m#1b) -- (#1*\bspWidth*1.5+0.7*\bspWidth,{\yshift{#1}-\bspArrowSize*0.5});
	\node (m#1) at ($(m#1a)!0.5!(m#1b)$) {};
}

\usepackage{xparse}

\newboolean{buildtikzpics}
\setboolean{buildtikzpics}{false} 

\newboolean{simpletrue}
\newboolean{simplefalse}
\setboolean{simpletrue}{true}
\setboolean{simplefalse}{false}

\usepackage[colorlinks,bookmarks=false,citecolor=blue,linkcolor=red,urlcolor=blue]{hyperref}
\usepackage{cleveref}

\begin{document}

\title{Formation of spinful dark excitons in Hubbard systems with magnetic superstructures}

\author{Constantin Meyer}
\affiliation{Institut f\"{u}r Theoretische Physik, Georg-August-Universit\"{a}t G\"{o}ttingen, Friedrich-Hund-Platz 1, D-37077 G\"{o}ttingen, Germany}
\author{Salvatore R. Manmana}
\affiliation{Institut f\"{u}r Theoretische Physik, Georg-August-Universit\"{a}t G\"{o}ttingen, Friedrich-Hund-Platz 1, D-37077 G\"{o}ttingen, Germany}
\affiliation{Fachbereich Physik, Philipps-Universit\"{a}t Marburg, Renthof 6, D-35032 Marburg, Germany}

\date{\today}

\begin{abstract}
The possibility to form excitons in photoilluminated correlated materials is central from fundamental and application oriented perspectives.  
In this paper we show how the interplay of electron-electron interactions and a magnetic superstructure leads to the formation of a peculiar spinful dark exciton, which can be detected in ARPES-type experiments and optical measurements. 
We study this by using matrix product states (MPS) to compute the time evolution of single-particle spectral functions and of the optical conductivity following an electron-hole excitation in a class of one-dimensional correlated band-insulators, simulated by Hubbard models with on-site interactions and alternating local magnetic fields. 
An excitation in only one specific spin direction leads to an additional band in the gap region of the spectral function only in the spin direction unaffected by the excitation and to an additional peak in the optical conductivity. 
As both is formed only after the excitation, this is interpreted as a dark exciton, which shows only in one spin direction. 
Recombination of the excitation happens on much longer time scales than the ones amenable to MPS.
We discuss implications for experimental studies in correlated insulator systems. 
\end{abstract}

\maketitle
\section{Introduction}

It is a central question if and how the interplay of strong electronic correlations and the excitation of a quantum many-body system by light leads to the formation of quasiparticles or of transient order, like charge density waves (CDW) or superconducting (SC) states \cite{RevModPhys.81.163,pumpprobe7,pumpprobe6,pumpprobe5,PhysRevLett.118.116402, 1367-2630-18-9-093028,Tao62,Rini2007,Hu2014,Ropers1,Fausti189,Mitrano2016,ncomms10459,PRLChromium,Schmitt1649,doublondynamics,Rohwer2011,Hellmann2012,Mathias2016,Stojchevska177,keunecke2020,keunecke2020direct,koehler2020formation_published,PhysRevB.101.180507}. 
Correlation effects are also discussed as possible sources for increasing the efficiency of photovoltaic devices\cite{Manousakis2010,Manousakis2019,Petocchi2019}, e.g., via the formation of multiple exciton generation due to impact ionization.
An important question is how excitons are formed, and which characteristics they possess in the presence of strong correlations \cite{Excitons2001,Jeckelmann2003,Dagotto2008,Al-Hassanieh2008}.
A hallmark of excitons is that Coulomb interaction leads to a binding energy between an electron excited to the conduction band and the remaining hole in the valence band\cite{kira_koch_2011}.
In Mott insulators, excitons lead to resonances, which can be in the gap region of the optical conductivity or within the Hubbard bands, and multiple exciton peaks are possible \cite{PhysRevLett.85.3910,Excitons2001,Jeckelmann2003}. 
More recently, the question has been studied how to identify excitons in ARPES-type measurements \cite{perfetto_melting_noneq_exc,wallauer_momentum_observation,dong_measurement_exciton,stefanucci_arpes_exciton,madeo_visualizing_dark_exc}.
For example in the theoretical studies \onlinecite{Bittner2020,christiansen2019,perfetto_melting_noneq_exc}, the binding energy of the exciton is shown to lead to a midgap signature in the corresponding spectral function.
Common to these studies is the presence of nearest-neighbor or longer-range Coulomb-interactions, which lead to the binding between the excited electron and the hole. 

In this paper, we investigate the formation of excitons in Hubbard systems with only on-site electron-electron interactions and an additional magnetic superstructure, but without longer-range Coulomb-interactions.
This is motivated by a one-dimensional toy model of manganites\cite{PhysRevB.97.235120} and the observation of orbital-selective Mott phases (OSMP)~\cite{Jacek_2019,Jacek_2020,Jacek_2020_2,Jacek_2021}. 
We treat the effect of a single, direct electron-hole excitation, in which an electron is assumed to be instantly excited over the gap without changing its momentum. 
Furthermore, we study the excitation in a single spin direction only, so that the electrons of the other spin direction are not touched by the incoming light. 
Such a spin-selective excitation can be studied using circularly polarized light and was shown in previous work to lead to the formation of spatially periodic charge-density or spin-density patterns\cite{koehler2020formation_published}. 
Here, we use time-dependent matrix product states (MPS)\cite{Schollwoeck201196,review_tdmrg} to study the time evolution of single-particle spectral functions\cite{kalthoff_2018,Costi1,Costi2,Costi3,Costi4,zawadzki_noneq_spectr,Zawadzki2020,zawadzki2020preprint}. 
In addition, we also compute the time-dependent optical conductivity\cite{PhysRevB.89.125123,PhysRevB.93.195144}.
In both quantities the photoexcitation leads to additional signals only in the \textit{opposite} spin direction than the one excited, which can be interpreted as a peculiar spinful dark exciton. 
We study the recombination process of the electron-hole excitation and find its time scale to be much longer than the ones amenable to MPS, which in the present case would correspond to $\sim 30$fs in pump-probe experiments. 

The remainder of this paper is structured as follows: 
In Sec.~\ref{sec:model} we introduce the models; in Sec.~\ref{sec:methods} we define the quantities studied by us, which are the time-dependent spectral functions and the time-dependent optical conductivity, and our numerical approach to compute them using MPS.
Our findings are presented in Sec.~\ref{sec:results}: 
In Sec.~\ref{sec:u} we discuss in detail the spectral functions in equilibrium and the effect of the Hubbard interaction $U$; in Sec.~\ref{sec:excitation} the effects of the electron-hole excitation immediately after its application are analyzed; in Sec.~\ref{sec:results_oc} we present our interpretation of the features as spinful dark excitons based on the findings on the spectral functions and the optical conductivity immediately after the excitation; in Sec.~\ref{sec:transient} we discuss the time evolution of the spectral function and of the optical conductivity, which show recombination of the excited electron-hole pair and stability of the excitonic feature on the time scales treated by us. 
Finally, a summary is provided in Sec.~\ref{sec:outlook}. 
The Appendix contains further aspects on the large $U$ behavior of the spectral functions and on the computation of the $k$-space properties in systems with a superstructure in the presence of open boundary conditions (OBC).

\section{Model}
\label{sec:model}

We study variants of the one-dimensional Hubbard model \cite{hubbard_original,Gutzwiller_HMorig,Hubbard_orig_Kanamori,book_hubbardmodel} with a magnetic superstructure, 
\begin{align}
\label{eq:model_pcmo}
	\hat{H} &= \hat{H}_{0} + \hat{H}_{U} = \hat{H}_{t_{\mathrm{h}}} + \hat{H}_{\Delta} + \hat{H}_{U}, \\
	\hat{H}_{t_{\mathrm{h}}} &= -t_{\mathrm{h}} \sum_{r,\sigma} \!\left(\hat{c}^{\dagger}_{\sigma, r+1}\hat{c}^{\vphantom{\dagger}}_{\sigma, r\vphantom{+1}} + \hat{c}^{\dagger}_{\sigma, r\vphantom{+1}}\hat{c}^{\vphantom{\dagger}}_{\sigma, r+1} \right)\!, \\
	\hat{H}_{U} &= U \sum_{r} \hat{n}^{\vphantom{\dagger}}_{\uparrow, r}\hat{n}^{\vphantom{\dagger}}_{\downarrow, r}, \\
	\hat{H}_{\Delta} &= \sum_{r} \Delta^{\vphantom{z}}_{r} S_{r}^{z} \quad\text{with}\quad S_{r}^{z} = \frac{1}{2}\left(\hat{n}^{\vphantom{\dagger}}_{\uparrow, r} - \hat{n}^{\vphantom{\dagger}}_{\downarrow, r}\right)\!. 
\end{align}
Here, $t_{\mathrm{h}}$ denotes the hopping amplitude, $U$ the on-site Coulomb repulsion, $\hat{c}^{\left(\dagger\right)}_{\sigma, r\vphantom{+1}}$ annihilates (creates) a particle of spin $\sigma$ at site $r$, $\hat{n}^{\vphantom{\dagger}}_{\sigma, r} = \hat{c}^{\dagger}_{\sigma, r} \hat{c}^{\vphantom{\dagger}}_{\sigma, r}$ is the particle density, and with $\Delta_{r}$ we describe an on-site Zeeman term of strength $\Delta$. 
We will consider two different setups, the first one with $\Delta_{r}$ alternating every two sites, $\left(\Delta, \Delta, -\Delta, -\Delta\right)$, and the second one with a site wise alternating $\Delta_{r}$, i.e. $\left(\Delta, -\Delta\right)$. 
We will use the superscripts $4\Delta$ and $2\Delta$ to distinguish between these two superlattices. 
As discussed in Ref.~\onlinecite{PhysRevB.97.235120}, these correspond to the ground states of a one-dimensional toy-manganite system at quarter and half filling, respectively. 
These ground states can be seen as crystals of polarons, in which immobile t$_{2g}$ electrons form the magnetic superlattice, which is experienced by itinerant e$_g$ electrons via Hund's coupling leading to the effective model \eqref{eq:model_pcmo}. 
Note that in Refs.~\onlinecite{Jacek_2019,Jacek_2020,Jacek_2020_2,Jacek_2021} similar structures were also found in OSMP states, which are obtained in ladder systems like BaFe$_2$Se$_3$, although the coupling of the conduction electrons there is realized via a Heisenberg exchange term rather than a Zeeman term.

Due to the four-site (two-site) unit cell, $\hat{H}^{4\Delta}_{0}$ ($\hat{H}^{2\Delta}_{0}$) exhibits four (two) clearly separated bands for finite $\Delta$ which we label by band indices $\nu = 1,2,3,4$ ($\nu = 1,2$).
In the non-interacting case $U=0$, $\hat{H}_{0}^{4\Delta}$ possesses the band structure\cite{PhysRevB.97.235120} 
\begin{align}
	\label{eq:model_disp_4}
	\varepsilon^{4\Delta}_{\nu}\!\left(k\right) &= s_{1,\nu} t_{\mathrm{h}} \sqrt{2 + \frac{\Delta^{2}}{4t_{\mathrm{h}}^{2}} + s_{2,\nu} 2\sqrt{\cos^{2}\!\left(2k\right) + \frac{\Delta^{2}}{4t_{\mathrm{h}}^{2}}}}, \\
	s &= 
	\begin{pmatrix}
		-1 & -1 & +1 & +1 \\
		+1 & -1 & -1 & +1
	\end{pmatrix}
\end{align}
with momenta
\begin{equation}
	k = -\frac{\pi}{4} + \frac{2\pi p}{4N}, \quad p \in \left\{0,\dots,N-1\right\}\!, 
\end{equation}
$N$ being the number of unit cells. 
Diagonalizing $\hat{H}^{2\Delta}_{0}$ gives
\begin{equation}
	\label{eq:model_disp_2}
	\varepsilon^{2\Delta}_{\nu}\!\left(k\right) = s_{\nu} t_{\mathrm{h}} \sqrt{ \frac{\Delta^{2}}{4t_{\mathrm{h}}^{2}} + 2\left(\cos\!\left(2k\right) + 1\right)} \,, \quad s_{\nu}=\left(-1\right)^{\nu} \, ,
\end{equation}
now for the momenta
\begin{equation}
	k = -\frac{\pi}{2} + \frac{2\pi p}{2N}, \quad p \in \left\{0,\dots,N-1\right\}\!. 
\end{equation}
The first Brillouin zone (BZ), thus, is $\left[-\pi/4, \dots, \pi/4\right[$ for $\hat{H}^{4\Delta}$ and $\left[-\pi/2, \dots, \pi/2\right[$ for $\hat{H}^{2\Delta}$, respectively.

These bands give a point of orientation to the identification of relaxation and recombination processes after an excitation of the model also in the presence of interactions.  

\section{MPS calculation of non-equilibrium dynamical response functions} 
\label{sec:methods}

We study the time evolution of the non-equilibrium single-particle spectral function
\begin{equation}
	\label{eq:spectral_function}
	\mathcal{A}^{<}_{\sigma}\!\left(k,\omega,t\right) = \mathcal{F}_{t^{\prime}\rightarrow \omega}\!\left[\left<\Psi\right| \hat{a}_{\sigma,k}^{\dagger}\!\left(t^{\prime}+t\right) \hat{a}_{\sigma,k}^{\phantom{\dagger}}\!\left(t\right)\left|\Psi\right>\right]
\end{equation}
and the time-dependent optical conductivity
\begin{equation}
	\sigma\!\left(\omega,t\right)=\sigma_{1}\!\left(\omega,t\right)+\mathrm{i}\sigma_{2}\!\left(\omega,t\right)=\frac{\mathcal{F}_{t^{\prime}\rightarrow \omega}\left[j_{\text{p}}\!\left(t^{\prime},t\right)\right]}{\mathrm{i}\!\left(\omega+\mathrm{i}\tilde{\eta}\right)L \mathcal{F}_{t^{\prime}\rightarrow \omega} \left[A_{\text{p}}\!\left(t^{\prime},t^{\vphantom{\prime}}\right)\right]} 
\end{equation}
before and after an excitation, with $\mathcal{F}$ denoting the Fourier transform. In the following, these expressions are explained and we discuss in some detail how these quantities are obtained with time-dependent MPS methods\cite{review_tdmrg}.
The time evolutions in the above expressions are computed using an MPS-implementation of the time-dependent variational principle (TDVP)\cite{tdvp_1, tdvp_2,review_tdmrg} in its two-site implementation from the SymMPS toolkit\cite{symmps}. 
If not mentioned otherwise, we use systems with OBC and our parameters are $L=32$ lattice sites, time step $\delta t =0.05$, and maximum MPS bond dimension $\chi_{\rm max} = 1000$. 

\subsection{Computation of time-dependent spectral functions with MPS}
\label{sec:methods_spectralfcts}

For non-interacting systems in equilibrium, the spectral function yields the system's band structure, 
\begin{equation}
	\mathcal{A}^{<,>}_{\sigma}\!\left(k,\omega\right) \propto \delta\!\left(\omega - \varepsilon^{<,>}_{\sigma}\!\left(k\right)\right).
\end{equation} 
Here, $<$ or $>$ denotes the lesser or greater spectral function, describing the occupied or unoccupied part of the band structure, respectively. 
In non-equilibrium setups the spectral function takes up an additional time dependence and its change in the course of time will characterize the response of the system to an excitation.
We will mainly study the time evolution of $\mathcal{A}^<(k,\omega)$, since it gives direct insight into the redistribution of populations of the occupied bands in the course of time, and also for the possible formation of new features after the excitation.

We compute the spectral function via Fourier transforming the corresponding real-space Green's function, e.g., for obtaining the lesser spectral function,
\begin{equation}
	\mathcal{G}^{<}_{\sigma}\!\left(r^{\vphantom{\prime}},r^{\prime},t^{\prime},t\right) = \left<\Psi\right| \hat{c}_{\sigma,r^{\vphantom{\prime}}}^{\dagger}\!\left(t^{\prime}+t\right) \hat{c}_{\sigma,r^{\prime}}^{\vphantom{\dagger}}\!\left(t\right)\left|\Psi\right> \, .
\label{eq:twotimegreen}
\end{equation}
Note that in equilibrium the state $\ket{\Psi}$ is an eigenstate of $\hat{H}$ (usually the ground state, if one is interested in properties at temperature $T=0$), so that for a time-independent Hamiltonian the time-dependence reduces to a single time variable $t'$, over which the Fourier transform to $\omega$-space is performed.
However, in a non-equilibrium situation, $\ket{\Psi}$ is either a state after some initial perturbation at time $t=t_0 \equiv 0$, or the Hamiltonian is modified at time $t_0$, e.g., in a quantum quench, so that it is not possible to reduce Eq.~\eqref{eq:twotimegreen} to a single time variable.
One performs the Fourier-transform to $\omega$-space by integrating over one of the two time variables (or a linear combination of them) using the second one as an explicit time-dependence of the resulting $\omega$-dependent quantity, leading to what we call a time-dependent spectral function \cite{pruschke_freericks,pruschke_freericks_erratum,review_pumpprobetheory,Costi1,Costi2,Costi3,Costi4,kalthoff_2018}.
In this approach, a freedom of choice is present for how to perform the Fourier transform in detail.
Equation~\eqref{eq:twotimegreen} is expressed using relative time coordinates. 
An alternative are Wigner coordinates, $t_{\text{rel}} = t^{\prime}$ and $t_{\text{ave}}=t+t^{\prime}/2$.
However, as discussed, e.g., in Ref.~\onlinecite{kalthoff_2018} both choices yield qualitatively similar results.
Here, we choose the relative coordinates, since for the computation of the expectation values in Eq.~\eqref{eq:twotimegreen} one needs to save fewer wave functions for the individual time steps needed in the computation.  

To obtain the $k$- and $\omega$-dependent spectral function, we first need to perform a transform of Eq.~\eqref{eq:twotimegreen} from real-space to (quasi-)momenta $k$, which for the OBC used by us is described in detail in Ref.~\onlinecite{koehler2020formation_published} and in App.~\ref{app:obc}.
With the annihilation (creation) operators $\hat{a}_{\sigma,k}^{(\dagger)}$ defined there we obtain from the data computed using Eq.~\eqref{eq:twotimegreen}:
\begin{align}
	\label{eq:tsfs_g}
	\mathcal{G}^{<}_{\sigma}\!\left(k,t^{\prime},t\right) &= \left<\Psi\right| \hat{a}_{\sigma,k}^{\dagger}\!\left(t^{\prime}+t\right) \hat{a}_{\sigma,k}^{\phantom{\dagger}}\!\left(t\right)\left|\Psi\right>\\
	\label{eq:tsfs}
	\mathcal{A}^{<}_{\sigma}\!\left(k,\omega,t\right) &= \mathcal{F}_{t^{\prime}\rightarrow \omega}\!\left[\mathcal{G}^{<}_{\sigma}\!\left(k,t^{\prime},t\right)\right] \, ,
\end{align}
where we follow Ref.~\onlinecite{PhysRevB.101.180507} and apply (up to factors $2 \pi$) the Fourier-transform
\begin{equation}
\label{eq:ft_t_w}
\mathcal{A}^{<}_{\sigma}\!\left(k,\omega,t\right) = 2 \mathrm{Re}\!\left[\int_{0}^{\infty} \mathrm{d}t^{\prime}\mathrm{e}^{-\mathrm{i}\omega t^{\prime}} \mathrm{e}^{-\eta t^{\prime}} \mathcal{G}^{<}_{\sigma}\!\left(k,t^{\prime},t\right)\right]\!.
\end{equation}
Here we have introduced a regularization via the damping factor $\eta$, which leads to a broadening of the signal, thus limiting the resolution in frequency. 
In the actual calculation, we discretize the integral. 
Due to this damping, the signals after a time $t^{\prime}_{\text{max}}$ decay to zero, so that we can restrict the simulations to this range. 
Likewise, the maximum time $t^{\prime}_{\text{max}}=8.0$ determines the smallest possible value for the frequencies that can be resolved\cite{NumRec}. 
To artificially increase the resolution of our data, we apply zero padding such that the input signal of the Fourier transform is of length $8\cdot 2^{\left\lceil\log_{2}\!\left(t^{\prime}_{\text{max}}/\delta t +1\right)\right\rceil}$. 
Note that in Eq.~\eqref{eq:ft_t_w} we perform the Fourier transform, implicitly assuming time-reversal symmetry, 
\begin{equation}
\label{eq:timerev}
	\mathcal{G}^{<}_{\sigma}\!\left(k,t^{\prime},t\right) = \mathcal{G}^{<*}_{\sigma}\!\left(k,-t^{\prime},t\right)\!.
\end{equation}

It is illustrative to consider the details of the computation needed to obtain the time-dependent spectral functions with MPS:
In the Schr\"{o}dinger picture we need to compute
\begin{align}
	\label{eq:step1}
	\left|\Psi\!\left(t^{\vphantom{\prime}}\right)\right> &= U\!\left(t^{\vphantom{\prime}},0\right)\left|\Psi\right>, \\
	\label{eq:step23}
	\left|\phi\!\left(t^{\prime}+t^{\vphantom{\prime}},t^{\vphantom{\prime}}\right)\right> &= U\!\left(t^{\prime}+t^{\vphantom{\prime}},t^{\vphantom{\prime}}\right) \hat{c}_{\sigma,r^{\prime}}^{\vphantom{\dagger}}\left|\Psi\!\left(t^{\vphantom{\prime}}\right)\right>, \quad \text{and} \\
	\label{eq:step4}
	\left|\Psi\!\left(t^{\prime}+t^{\vphantom{\prime}}\right)\right> &= U\!\left(t^{\prime}+t^{\vphantom{\prime}},t^{\vphantom{\prime}}\right) \left|\Psi\!\left(t^{\vphantom{\prime}}\right)\right> = U\!\left(t^{\prime}+t^{\vphantom{\prime}},0\right) \left|\Psi\right>, 
\end{align}
where $U\!\left(t^{\vphantom{\prime}}_{2},t^{\vphantom{\prime}}_{1}\right)$ is the time evolution operator for a real time evolution from time $t^{\vphantom{\prime}}_{1}$ to $t^{\vphantom{\prime}}_{2}$ and the operator application in Eq.~\eqref{eq:step23} is conducted variationally\cite{review_tdmrg}. 
The real space and two-time-dependent lesser Green's function is then computed by the overlap
\begin{equation}
	\mathcal{G}^{<}_{\sigma}\!\left(r^{\vphantom{\prime}},r^{\prime},t^{\prime},t\right) = \left<\Psi\!\left(t^{\prime}+t^{\vphantom{\prime}}\right)\right| \hat{c}_{\sigma,r^{\vphantom{\prime}}}^{\dagger}\left|\phi\!\left(t^{\prime}+t^{\vphantom{\prime}},t^{\vphantom{\prime}}\right)\right>.
\end{equation}
Note that this procedure works for time-independent and time-dependent systems alike.
 
Note that the $k$-dependence in Eqs.~\eqref{eq:tsfs_g} and~\eqref{eq:tsfs} corresponds to the extended zone scheme. 
Both the Greens functions, as well as the spectral functions can be transformed to the reduced zone scheme, i.e. the first BZ, see Eqs.~\eqref{eq:model_fold_back_4} or \eqref{eq:model_fold_back_2}, respectively leading to the band index $\nu$. 
Unless explicitly mentioned, we will show the sum over the band index $\nu$. 

For the sake of comparison with exact diagonalization at equilibrium, we also show results obtained from the Lehmann representation\cite{mah00}, 
\begin{equation}
	\mathcal{A}^{<}_{\sigma}\!\left(k,\omega\right) = \sum_{n}\left|\left<n\right|\hat{a}^{\vphantom{\dagger}}_{\sigma,k}\left|\mathrm{GS}\right>\right|^{\!2} \delta\!\left(E_{0}-E_{n}-\omega\right)
\end{equation}
in the extended zone scheme and 
\begin{equation}
	\label{eq:lehmann}
	\mathcal{A}^{<}_{\sigma}\!\left(k,\omega\right) = \sum_{n,\nu}\left|\left<n\right|\hat{a}^{\vphantom{\dagger}}_{\sigma,\nu,k}\left|\mathrm{GS}\right>\right|^{\!2} \delta\!\left(E_{0}-E_{n}-\omega\right) 
\end{equation}
in the first  BZ. 
Here, $\left|n\right>$ describe the system's eigenstates, $E_{n}$ their corresponding eigenenergies, $E_{0}$ is the ground state energy, and $\delta$ the Dirac delta function. 
We will consider only a single unit cell, so that Eq.~\eqref{eq:lehmann} equals its $k$-independent form, 
\begin{equation}
	\label{eq:lehmann_calc}
	\begin{split}
		\mathcal{A}^{<}_{\sigma}\!\left(\omega\right) &= \sum_{n,r}\left|\left<n\right|\hat{c}^{\vphantom{\dagger}}_{\sigma,r}\left|\mathrm{GS}\right>\right|^{\!2} \delta\!\left(E_{0}-E_{n}-\omega\right) \\
		&= \sum_{k}\mathcal{A}^{<}_{\sigma}\!\left(k,\omega\right). \\
	\end{split} 
\end{equation}

For the greater Greens functions similarly one has
\begin{equation}
	\mathcal{G}^{>}_{\sigma}\!\left(r^{\vphantom{\prime}},r^{\prime},t^{\prime},t\right) = \left<\Psi\right| \hat{c}_{\sigma,r^{\vphantom{\prime}}}^{\vphantom{\dagger}}\!\left(t\right) \hat{c}_{\sigma,r^{\prime}}^{\dagger}\!\left(t^{\prime}+t\right)\left|\Psi\right>
\end{equation}
leading to
\begin{equation}
	\label{eq:lehmann_g}
	\mathcal{A}^{>}_{\sigma}\!\left(k,\omega\right) = \sum_{n,\nu}\left|\left<n\right|\hat{a}^{\dagger}_{\sigma,\nu,k}\left|\mathrm{GS}\right>\right|^{\!2} \delta\!\left(E_{n}-E_{0}-\omega\right) \,. 
\end{equation}

\subsection{Computation of the optical conductivity}
\label{sec:methods_oc}

We apply the scheme introduced in Refs.~\onlinecite{PhysRevB.93.195144,PhysRevB.89.125123}.
In this way one does not need to compute the costly current-current correlation functions, instead it suffices to compute the response-current following a probe-pulse, which we model via Peierls substitution\cite{Peierls1933,Mentink2015,PhysRevB.88.075135} 
\begin{equation}
	\hat{H}_{t_{\mathrm{h}}} \rightarrow \hat{H}_{t_{\mathrm{h}},\mathrm{p}} = -t_{\mathrm{h}} \sum_{r,\sigma} \!\left(\mathrm{e}^{\mathrm{i}A_{\text{p}}\!\left(t^{\prime},t\right)}\hat{c}^{\dagger}_{\sigma, r+1}\hat{c}^{\vphantom{\dagger}}_{\sigma, r\vphantom{+1}} + \text{h.c.}\right)
	\label{eq:peierls}
\end{equation}
and which we assume to be Gaussian shaped,
\begin{equation}
	A_{\text{p}}\!\left(t^{\prime},t\right) = A_{0, \text{p}} \mathrm{e}^{-\frac{\left(t+t^{\prime}-\Delta t\right)^{2}}{2\tau^{2}}}\cos\!\left(\omega_{\text{p}}\!\left(t+t^{\prime}-\Delta t\right)\right)\, .
\end{equation}
The value of the shift $\Delta t$ is chosen to be small, so that the peak of the pulse is very close to the instant $t$, at which we want to evaluate the optical conductivity, but such that the cut-off of the pulse at $t$ is negligible.
This leads to an explicit time dependence in the current operator ($t^{\prime} \geq 0$), 
\begin{equation}
	\hat{\jmath}\!\left(t^{\prime},t\right) =  -\mathrm{i}t_{\mathrm{h}}\sum_{r,\sigma}\!\left(\mathrm{e}^{\mathrm{i}A_{\text{p}}\!\left(t^{\prime},t\right)}\hat{c}^{\dagger}_{\sigma, r+1}\hat{c}^{\vphantom{\dagger}}_{\sigma, r\vphantom{+1}} - \text{h.c.}  \right)\, ,
\end{equation} 
and in the time evolution operator $U_p(t_1,t_2)$, in which $\hat{H}_{t_{\mathrm{h}}}$ is replaced by $\hat{H}_{t_{\mathrm{h}},\mathrm{p}}$. 
Note that for spin-resolved results, we omit the sum over $\sigma$ and compute two current operators for $\sigma = \uparrow$ and $\sigma = \downarrow$ individually. 
Other than this the scheme works analogously. 
The response current $j_{\rm p}$ is the difference to the current present in the system at time $t$ without the probe pulse, 
\begin{equation}
	\label{eq:methods_oc_j}
	\begin{split}
	j_{\text{p}}\!\left(t^{\prime},t\right) = {} &\left<\Psi\!\left(t\right)\right|U_{\text{p}}\!\left(t,t^{\prime}+t\right)\hat{\jmath}\!\left(t^{\prime},t\right)U_{\text{p}}\!\left(t^{\prime}+t,t\right)\left|\Psi\!\left(t\right)\right> \\
	 &-\left<\Psi\!\left(t\right)\right|U\!\left(t,t^{\prime}+t\right)\hat{\jmath}\!\left(t^{\prime},t\right)U\!\left(t^{\prime}+t,t\right)\left|\Psi\!\left(t\right)\right> \, .
	\end{split}
\end{equation}
This gives the expression
\begin{equation}
	\sigma\!\left(\omega,t\right) \equiv \sigma_{1}\!\left(\omega,t\right)+\mathrm{i}\sigma_{2}\!\left(\omega,t\right)=\frac{j_{\text{p}}\!\left(\omega,t\right)}{\mathrm{i}\!\left(\omega+\mathrm{i}\tilde{\eta}\right)LA_{\text{p}}\!\left(\omega,t\right)} \, ,
\end{equation}
where 
\begin{align}
	\label{eq:methods_oc_ft_a}
	A_{\text{p}}\!\left(\omega,t\right) &= \mathcal{F}_{t^{\prime}\rightarrow \omega}\!\left[A_{\text{p}}\!\left(t^{\prime},t\right)\right] \quad \text{and} \\ 
	\label{eq:methods_oc_ft_j}
	j_{\text{p}}\!\left(\omega,t\right) &= \mathcal{F}_{t^{\prime}\rightarrow \omega}\!\left[j_{\text{p}}\!\left(t^{\prime},t\right)\right]
\end{align} 
describe the Fourier transforms of the time-dependent vector potential of the probe pulse and of the current, respectively.

To compute the Fourier transforms, we proceed as described after Eq.~\eqref{eq:ft_t_w}. 
Note, however, that in this case we do not assume time-reversal symmetry, so that we take track of the real and the imaginary part and 
we include a damping factor $\tilde{\eta}$ in both cases. 
(In general we take $\tilde{\eta} \neq \eta$.)
Furthermore, we apply zero padding enhancing our data by $16$ times its length with zeros. 
We ensure that we are in the linear-response regime by choosing amplitude and width of the probe pulse small enough. 
Unless stated otherwise, we will work with probe pulses with $\Delta t =0.25$, $\tau = 0.05$, $A_{0} = 0.5$ and $\omega_{\text{p}}=2.38$. 
Furthermore, we choose $\tilde{\eta}=0.1$, $t^{\prime} \in \left[0, 25\right]$, and use time steps of $\delta t = 0.05$. 
Note that for these computations we took $\chi_{\text{max}}=500$. 
Regarding the accuracy at small frequencies we use the following estimate: For the minimum frequency resolvable we take $\omega_{\text{min}} = 2\pi/t^{\prime}_{\text{max}} \sim \pi / 10 \sim 0.3$. 
In addition, we need to consider the broadening $\tilde{\eta}$ which adds to the above value. 
In a conservative estimate, we multiply this value by two such that we consider results at frequencies larger than $\omega_{\text{min}} \sim 1$. 

\section{Results}
\label{sec:results}

\begin{figure*}
	\includegraphics[width=\textwidth]{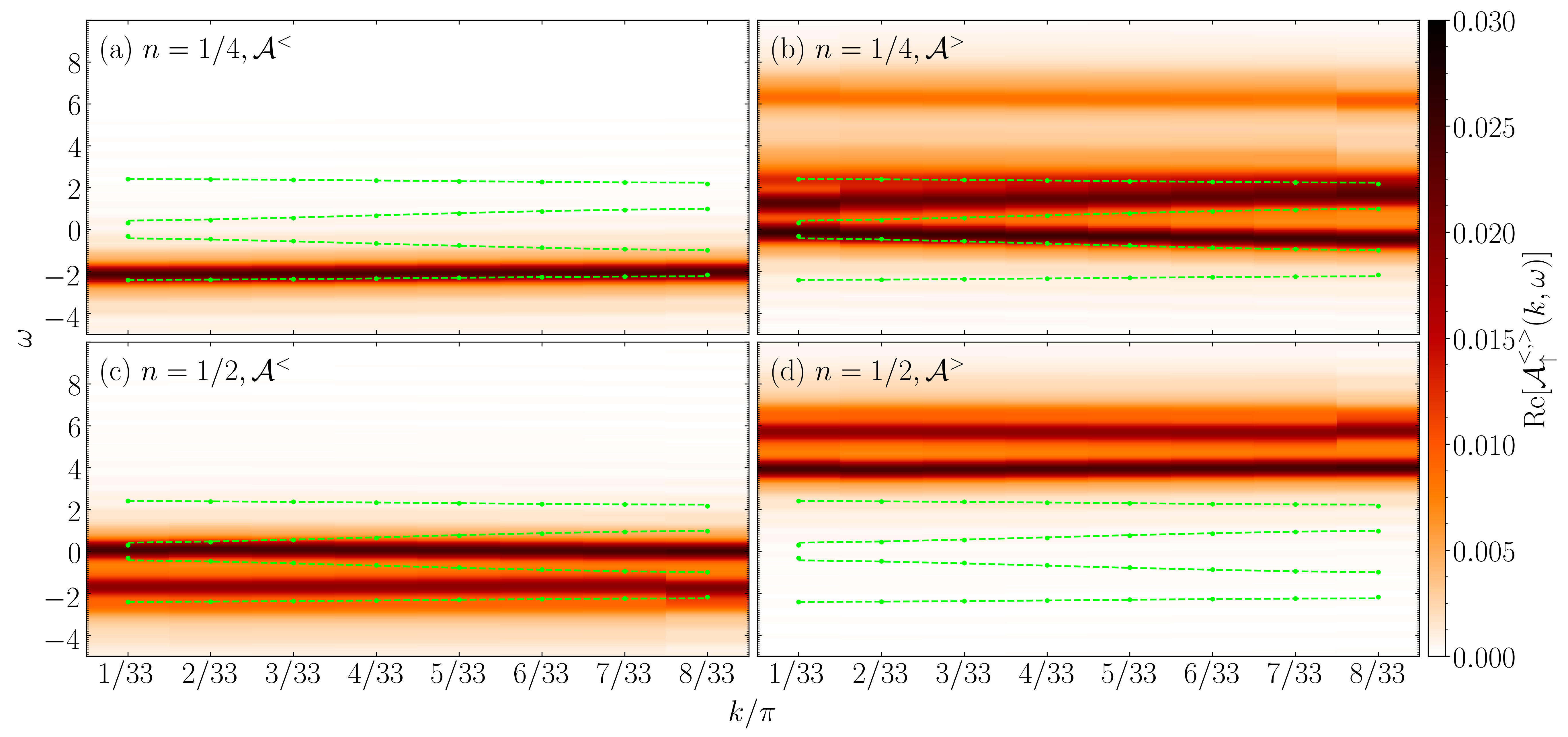}
	\caption{Single particle spectral functions $\mathcal{A}^{<}\!\left(k,\omega\right)$ and $\mathcal{A}^{>}\!\left(k,\omega\right)$ in equilibrium for $\hat{H}^{4\Delta}$ with $U/t_{\mathrm{h}} = 4$, $L = 32, \Delta/t_{\mathrm{h}} = 2$ obtained with MPS for OBC in the first BZ at quarter, (a) and (b), and half filling, (c) and (d), respectively. 
		Only the results for the $\uparrow$-direction are shown, as the results for the $\downarrow$-direction are identical. 
		The green dashed lines show the band structure of the non-interacting system 
		calculated with PBC, the green dots correspond to the calculation with OBC. 
		The discontinuities at the edges of the first BZ ($k = \pi/33$ and $k=8\pi/33$) are due to OBC. ($\chi_{\text{max}} = 500$.)
	}
	\label{fig:eff_u}
\end{figure*}
We now discuss our results first in equilibrium at temperature $T=0$, and second directly after an electron-hole type excitation, which we apply only to one spin direction, which serves two purposes: i) this allows us to study in more detail the interplay of the excitation and interaction effects between the electrons by analyzing the subsequent evolution in the two spin directions separately; ii) this procedure is reminiscent of spin-selective excitations, which can be realized, e.g., in spin-polarized ARPES experiments, typically by circularly polarized light\cite{Bruno_moke}.
A similar spin-selective photoexcitation of such models has been studied in Ref.~\onlinecite{koehler2020formation_published}, where the formation of periodic patterns in real space is reported.
Here, in contrast, we are interested in the effect on the dynamical quantities, as further detailed below.
After investigating the stronger correlation effects at half filling in the model $\hat{H}^{4\Delta}$, we will see to which extend they are also realized in $\hat{H}^{2\Delta}$.

\subsection{Effect of $U$ on the spectral functions}
\label{sec:u}

For strongly interacting systems the distribution of the spectral weight cannot be associated to bands in a stricter sense due to the strong scattering between the electrons. 
Nevertheless, it is a good point of reference for our systems and we will compare the results for the spectral function at finite $U$ with the non-interacting band structure.

We first investigate $\hat{H}^{4\Delta}$ restricting ourselves to half or quarter filling, such that in the non-interacting system either the lowest or the lowest two bands are fully occupied, realizing a band-insulator in both cases. 
As discussed before, at quarter filling this corresponds to the ground state of the toy-manganite system of Ref.~\onlinecite{PhysRevB.97.235120}. 
When going to half filling, due to the larger number of interacting particles, we expect stronger correlation effects at finite $U$. 

Figure~\ref{fig:eff_u} displays the lesser and greater spectral functions, $\mathcal{A}^{<}\!\left(k,\omega\right)$ and $\mathcal{A}^{>}\!\left(k,\omega\right)$, folded back to the first BZ, for an $L=32$-site system with $U/t_{\mathrm{h}} = 4$, and $\Delta/t_{\mathrm{h}} = 2$, which are approximately the ab-initio values of Ref.~\onlinecite{PhysRevB.97.235120}. 
The dashed lines show the position of the non-interacting bands \eqref{eq:model_disp_4} and serve as a reference.
At quarter filling, for $\mathcal{A}^{<}\left(k,\omega\right)$, c.f. Fig.~\ref{fig:eff_u}(a), 
$U$ plays only a minor role:
the results resemble the non-interacting case, apart from a slight shift to higher frequencies and a weak flattening of the occupied band.
In contrast, $\mathcal{A}^{>}\!\left(k,\omega\right)$ shown in Fig.~\ref{fig:eff_u}(b), deviates strongly from the non-interacting band structure. 
While the second band is rather well reproduced, the third band shows a considerable shift.
In addition, a new feature at $\omega \approx 6$ is obtained, which is approximately at an energy of $U/\hopping$ higher than the third band. 
The fourth band in the non-interacting case appears to have split into two. 

At half filling, the effect of the interaction is substantial also for $\mathcal{A}^{<}\left(k,\omega\right)$. 
As seen in Fig.~\ref{fig:eff_u}(c), an additional band below the lowest $\nu = 1$ band of the non-interacting system is realized.
Due to the particle-hole symmetry of the system at half filling the greater spectral function $\mathcal{A}^{>}\!\left(k,\omega\right)$ reflects the behavior of $\mathcal{A}^{<}\!\left(k,\omega\right)$, but mirrored at $U/2t_{\mathrm{h}}$, see Fig.~\ref{fig:eff_u}(d).
This can be interpreted as two Hubbard bands (a filled lower Hubbard band and an empty upper Hubbard band), which, however, due to the finite value of $\Delta$, possess a further fine structure.
The system hence is a 'hybrid' of a band- and of a strongly correlated insulator, with a stronger reminiscence to Mott-Hubbard insulators\cite{book_gebhard} due to the existence of the two symmetric Hubbard bands. 
In the following we will therefore refer to our system as correlated band insulator.
The features in all cases shown so far appear to be rather flat, i.e. the dispersion depends only weakly on $k$. 

\begin{figure}[b]
	\includegraphics[width=0.48\textwidth]{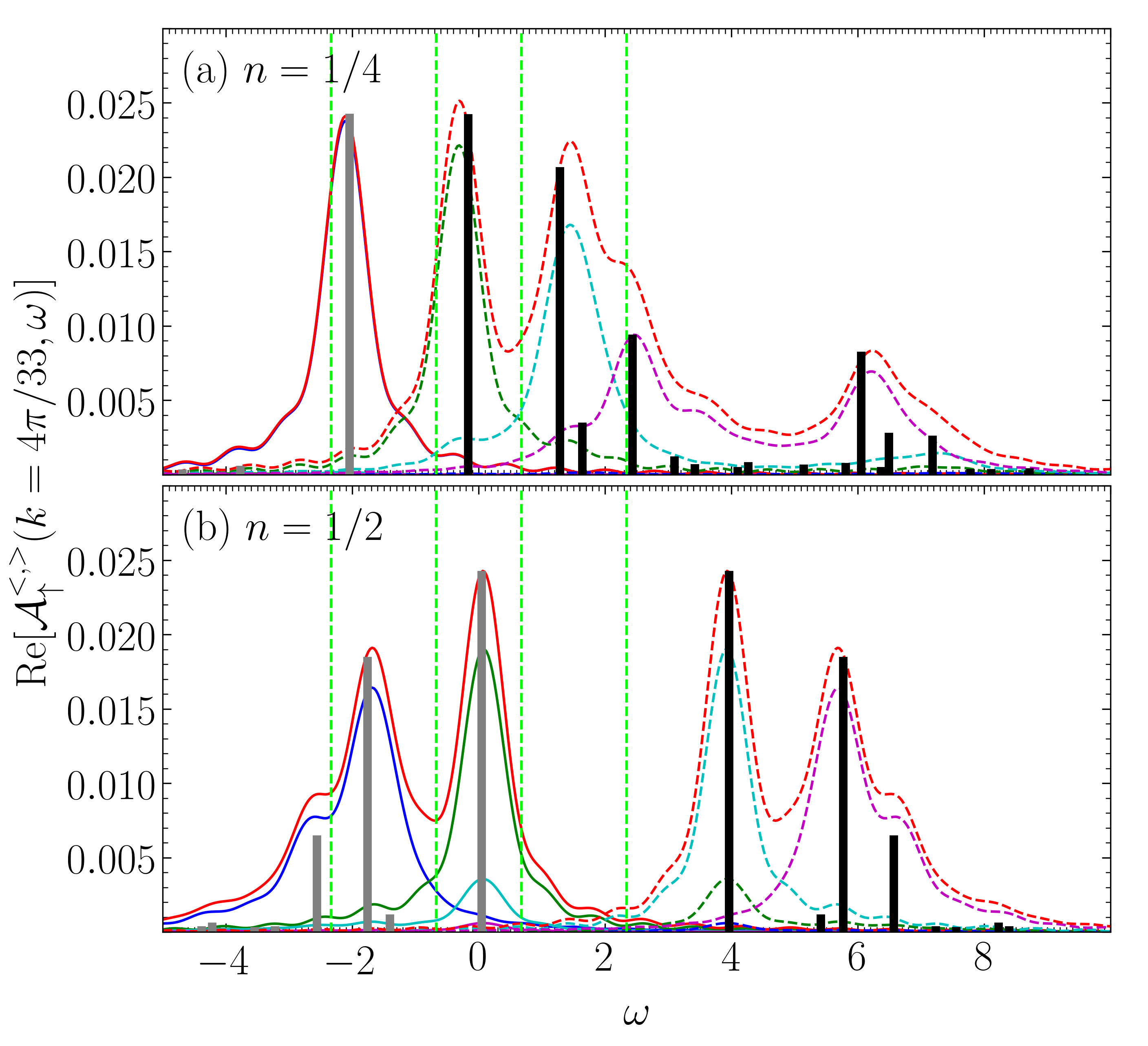}
	\caption{Cross section of the data shown in Fig.~\ref{fig:eff_u} for (a) quarter and (b) half filling. Solid lines: $\mathcal{A}^{<}_{\uparrow}\!\left(k = 4\pi/33,\omega\right)$; dashed lines: $\mathcal{A}^{>}_{\uparrow}\!\left(k = 4\pi/33,\omega\right)$.
		The total spectral function is shown in red, the contribution from the four band indices $\nu=1,2,3,4$ are given in blue, green, cyan, and magenta, respectively. 
		Vertical dashed light green lines: peak positions in the non-interacting case obtained for PBC. 
		Vertical gray (black) bars: excitation energies for $\mathcal{A}^{<}_{\uparrow}$ ($\mathcal{A}^{>}_{\uparrow}$) as obtained from Eq. \eqref{eq:lehmann_calc} for one unit cell, i.e. $L=4$. The heights of the bars correspond to their respective weights. 
		All bars have been scaled such that in each cross section the largest bar (irrespective of belonging to $\mathcal{A}^{<}_{\uparrow}$ or $\mathcal{A}^{>}_{\uparrow}$) takes a value of $0.8$ times the plot's maximum range in $y$-direction. ($\chi_{\text{max}} = 500$.)
	}
	\label{fig:crosssection}
\end{figure}

In Fig.~\ref{fig:crosssection} we show results for the cross section of the spectral function at $k = 4\pi/33 \approx \pi/8$, which is at the center of half of the first BZ.
We complement these MPS results by full diagonalization (FD) of one unit cell, i.e. $L=4$ for OBC, for which we obtain the spectral function via the Lehmann representation~\eqref{eq:lehmann}.
This is useful, since we have a comparably large broadening $\eta \approx 0.1$ in the MPS results. 
Note that one can substantially increase the resolution by going to larger systems\cite{review_tdmrg} or by applying Chebyshev expansions\cite{weisse_kernel,Holzner_chebyshev}.
However, further below we will compute the spectral functions out-of-equilibrium, which is computationally a substantially more costly task, restricting ourselves to treat small systems with a finite resolution.
In order to discuss the results on the same footing, we use the same set up also for the discussion of the equilibrium properties. 
We find a good agreement between our MPS and the FD results.
In particular, the positions of the weights of the FD results are in nearly perfect agreement with the peak positions of the MPS data for larger systems, indicating that this structure remains when going to large systems.  

\begin{figure}
	\centering
	\includegraphics[width=0.48\textwidth]{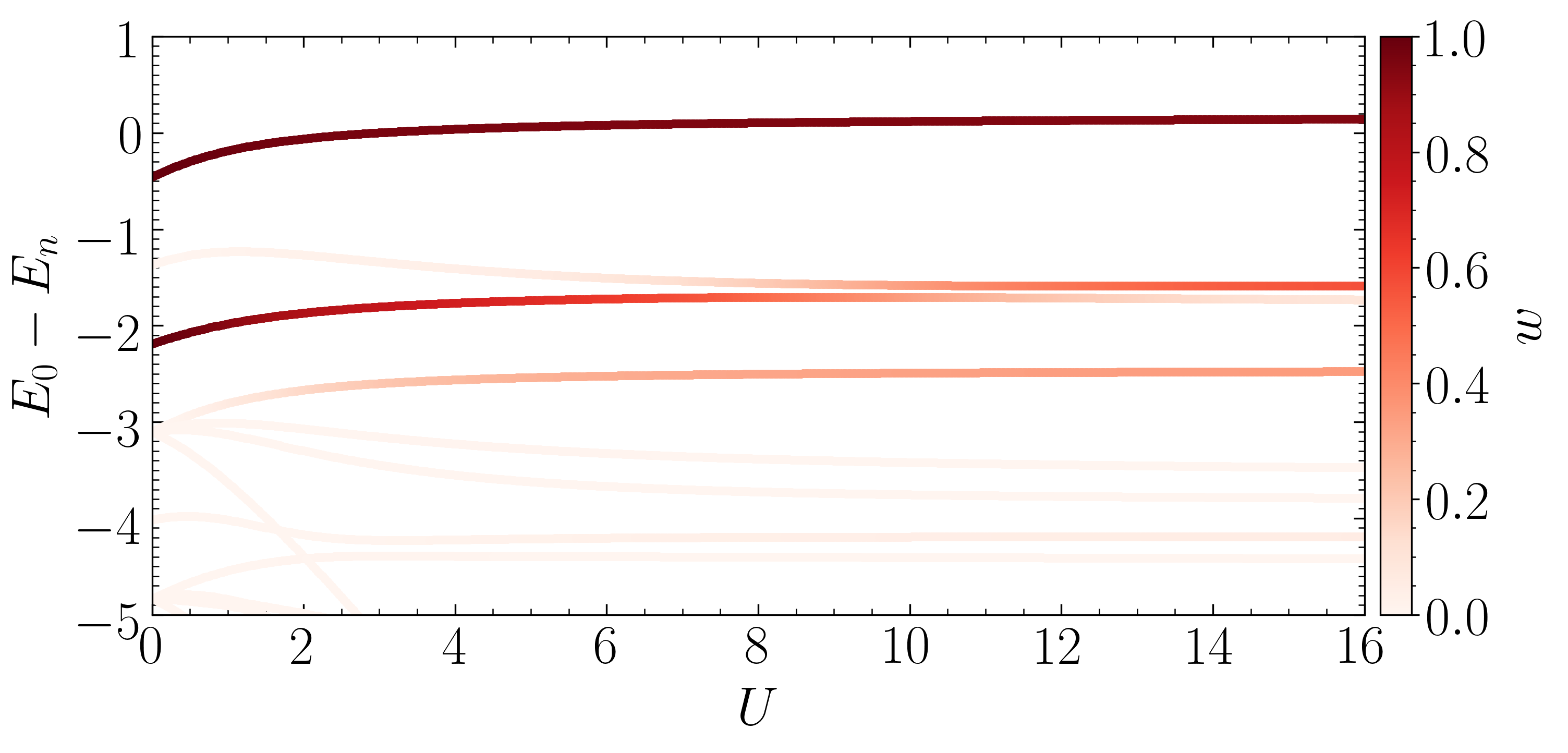}
	\caption{$\mathcal{A}^{<}_{\uparrow}\!\left(k,\omega\right)$ as a function of $U$ obtained from FD of a single unit cell of $\hat{H}^{4 \Delta}$ with OBC according to Eq.~\eqref{eq:lehmann_calc} for the case of half filling. The intensity of the line color indicates the weight $w$ of the respective peaks $w_{n}\!\left(U\right) = \sum_{r} \left|\left<n\right|\hat{c}_{\uparrow,r}\left|\mathrm{GS}\right>\!\left(U\right)\right|^{2}$. At $U=0$ the non-interacting band structure is reproduced. 
	The figure was cut at $E_{0}-E_{n}=-5$ below which all peaks occur with $w \ll 0.2$.
	}
	\label{fig:u_lehmann_l}
\end{figure}

Thus, we use the FD results in Fig.~\ref{fig:u_lehmann_l} to further analyze the influence of $U$ on the band structure.
We focus on the half filled case. 
Further details and results for larger values of $U$ are discussed in App.~\ref{app:u}.
Fig.~\ref{fig:u_lehmann_l} shows that already at small or moderate values of $U$ additional weights appear in the spectral function, which cannot be traced back to the non-interacting band structure.
For example, we find that a weak additional structure is formed at $\omega \approx -3$, which appears to split up when increasing $U$.
\begin{figure*}
	\includegraphics[width=\textwidth]{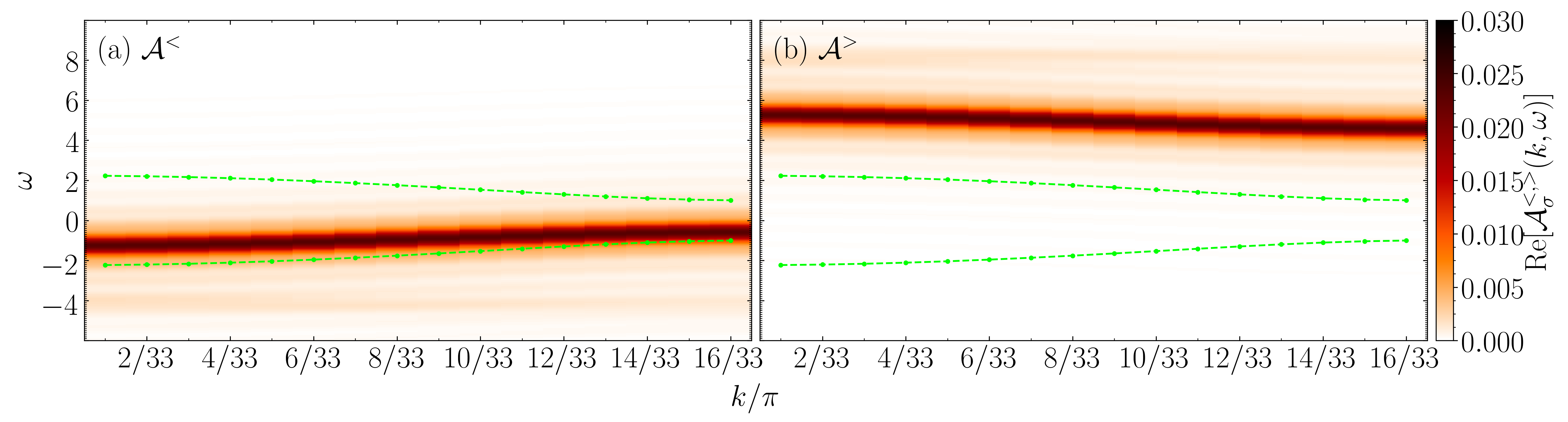}
	\caption{Single particle spectral functions $\mathcal{A}^{<}\!\left(k,\omega\right)$ (a) and $\mathcal{A}^{>}\!\left(k,\omega\right)$ (b) in equilibrium for $\hat{H}^{2\Delta}$ with $U/t_{\mathrm{h}} = 4$, $L = 32, \Delta/t_{\mathrm{h}} = 2$ obtained with MPS for OBC at half filling. 
		Only the results for the $\uparrow$-direction are shown, as the results for the $\downarrow$-direction are identical. 
		The green dashed lines show the band structure of the non-interacting system 
		calculated with PBC, the green dots correspond to the calculation with OBC. 
		In contrast to Fig.~\ref{fig:eff_u} there is no discontinuity at the edges of the first BZ. 
	}
	\label{fig:eff_u_2pcmo}
\end{figure*}
Also, we observe that both the first (around $\omega \approx -2$) and the second band (around $\omega \approx 0$) are subject to a renormalization with growing $U$. 
However, while the position of the second band appears to saturate at a value of $\omega \approx 0$, a new structure emerges in the vicinity of the first band, which for $U \gtrsim 10$ then becomes the dominant feature, apparently taking over the weight from the original first band. 
Starting at $U \approx 2$ we find that one of the new signals beneath the original first band at $\omega \approx -2.5$ substantially gains weight with increasing $U$. 
This explains the apparent band splitting of the $\nu=1$ band observed in Fig.~\ref{fig:crosssection}(b).
We also encounter further very weak signals at $\omega \approx -4$, which show a trace in Fig.~\ref{fig:crosssection}(b).
Thus, at $U=4$, $\mathcal{A}^{<}\!\left(k,\omega\right)$ displayed in Figs.~\ref{fig:eff_u}(c) and \ref{fig:crosssection}(b), predominantly possesses two renormalized bands, which stem from the non-interacting band structure, and additional features, which are correlation induced, two of them with significant weight. 
The total spectral function, hence, is composed of two Hubbard bands with an additional peak structure. 

\begin{figure}[b]
	\includegraphics[width=0.48\textwidth]{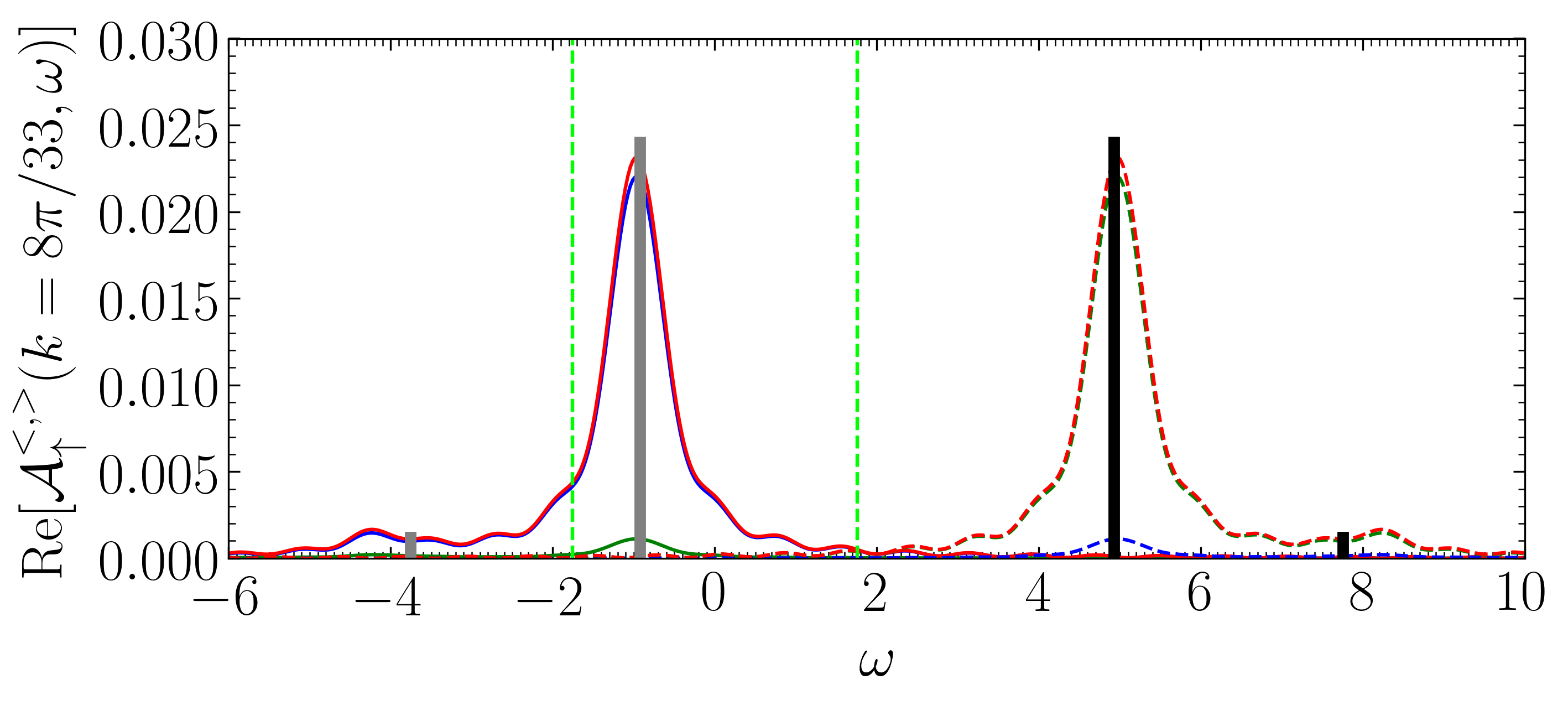}
	\caption{Cross section of the data shown in Fig.~\ref{fig:eff_u_2pcmo} for half filling. Solid lines: $\mathcal{A}^{<}_{\uparrow}\!\left(k = 8\pi/33,\omega\right)$; dashed lines: $\mathcal{A}^{>}_{\uparrow}\!\left(k = 8\pi/33,\omega\right)$.
		The total spectral function is shown in red, the contribution from the two band indices $\nu=1,2$ are given in blue and green, respectively. 
		Vertical dashed light green lines: peak positions in the non-interacting case obtained for PBC. 
		Vertical gray (black) bars: excitation energies for $\mathcal{A}^{<}_{\uparrow}$ ($\mathcal{A}^{>}_{\uparrow}$) as obtained from Eq. \eqref{eq:lehmann_calc} for one unit cell, i.e. $L=2$. The heights of the bars correspond to their respective weights. 
		All bars have been scaled such that in each cross section the largest bar (irrespective of belonging to $\mathcal{A}^{<}_{\uparrow}$ or $\mathcal{A}^{>}_{\uparrow}$) takes a value of $0.8$ times the plot's maximum range in $y$-direction.}
	\label{fig:crosssection_2pcmo}
\end{figure}

\begin{figure}[b!]
	\centering
	\includegraphics[width=0.48\textwidth]{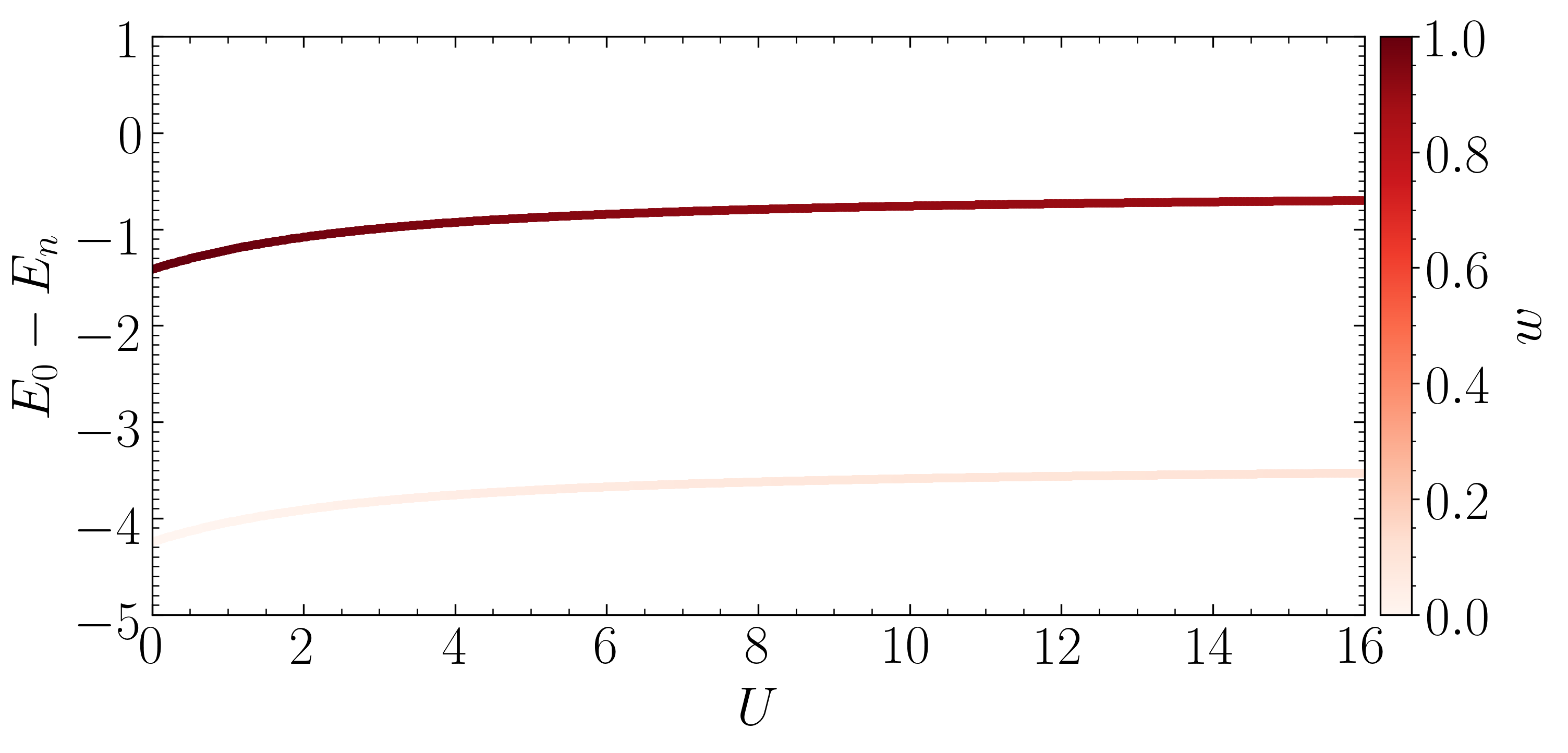}
	\caption{$\mathcal{A}^{<}_{\uparrow}\!\left(k,\omega\right)$ as a function of $U$ obtained from FD of a single unit cell of $\hat{H}^{2 \Delta}$ with OBC according to Eq.~\eqref{eq:lehmann_calc} for the case of half filling. The intensity of the line color indicates the weight $w$ of the respective peaks $w_{n}\!\left(U\right) = \sum_{r} \left|\left<n\right|\hat{c}_{\uparrow,r}\left|\mathrm{GS}\right>\!\left(U\right)\right|^{2}$. At $U=0$ the non-interacting band structure is reproduced. 
	}
	\label{fig:u_lehmann_l_2pcmo}
\end{figure}

To test if these features are realized also in other Hubbard systems with a magnetic superlattice, we now consider the same quantitites for $\hat{H}^{2\Delta}$ at half filling. 
Figure~\ref{fig:eff_u_2pcmo} shows $\mathcal{A}^{<}\!\left(k,\omega\right)$ and $\mathcal{A}^{>}\!\left(k,\omega\right)$ in the first BZ for a system with $L=32$ sites at $U/t_{\mathrm{h}} = 4$, and $\Delta/t_{\mathrm{h}} = 2$ in analogy to Fig.~\ref{fig:eff_u}. 
Compared to $\hat{H}^{4\Delta}$ we find a stronger dispersion, which decreases for larger $U$. 
Similar to the behavior found for $\hat{H}^{4 \Delta}$, a renormalization of the band structure is obtained, as well as the formation of an additional signal below and above the dominant contribution for $\mathcal{A}^{<}\!\left(k,\omega\right)$ in Fig.~\ref{fig:eff_u_2pcmo}(a) and $\mathcal{A}^{>}\!\left(k,\omega\right)$ in Fig.~\ref{fig:eff_u_2pcmo}(b), respectively. 
This is further illustrated in Fig.~\ref{fig:crosssection_2pcmo}, which shows the cross sections at $k = 4\pi/33 \approx \pi/8$ and additional FD results for a single unit cell.
We find $\mathcal{A}^{<}\!\left(k,\omega\right)$ to be almost entirely determined by contributions corresponding to first band index $\nu = 1$, while the same applies to  $\mathcal{A}^{>}\!\left(k,\omega\right)$ for $\nu=2$.
Note that the stronger dispersion makes the agreement between the MPS and the FD less good compared to what is seen for $\hat{H}^{4\Delta}$. 
Due to the smaller unit cell of $L=2$, the spectrum in the Lehmann representation has fewer features than for $\hat{H}^{4\Delta}$ consisting only of two contributions as seen in Fig.~\ref{fig:crosssection_2pcmo}.
From Fig.~\ref{fig:u_lehmann_l_2pcmo}, where we have again analyzed the influence of $U$ on a system composed of one unit cell analogously to Fig.~\ref{fig:u_lehmann_l}, we learn that with growing $U$ both of these signals are only subject to a slight renormalization. 

\subsection{Electron-hole-like excitation}
\label{sec:excitation}

\begin{figure*}
	\includegraphics[width=\textwidth]{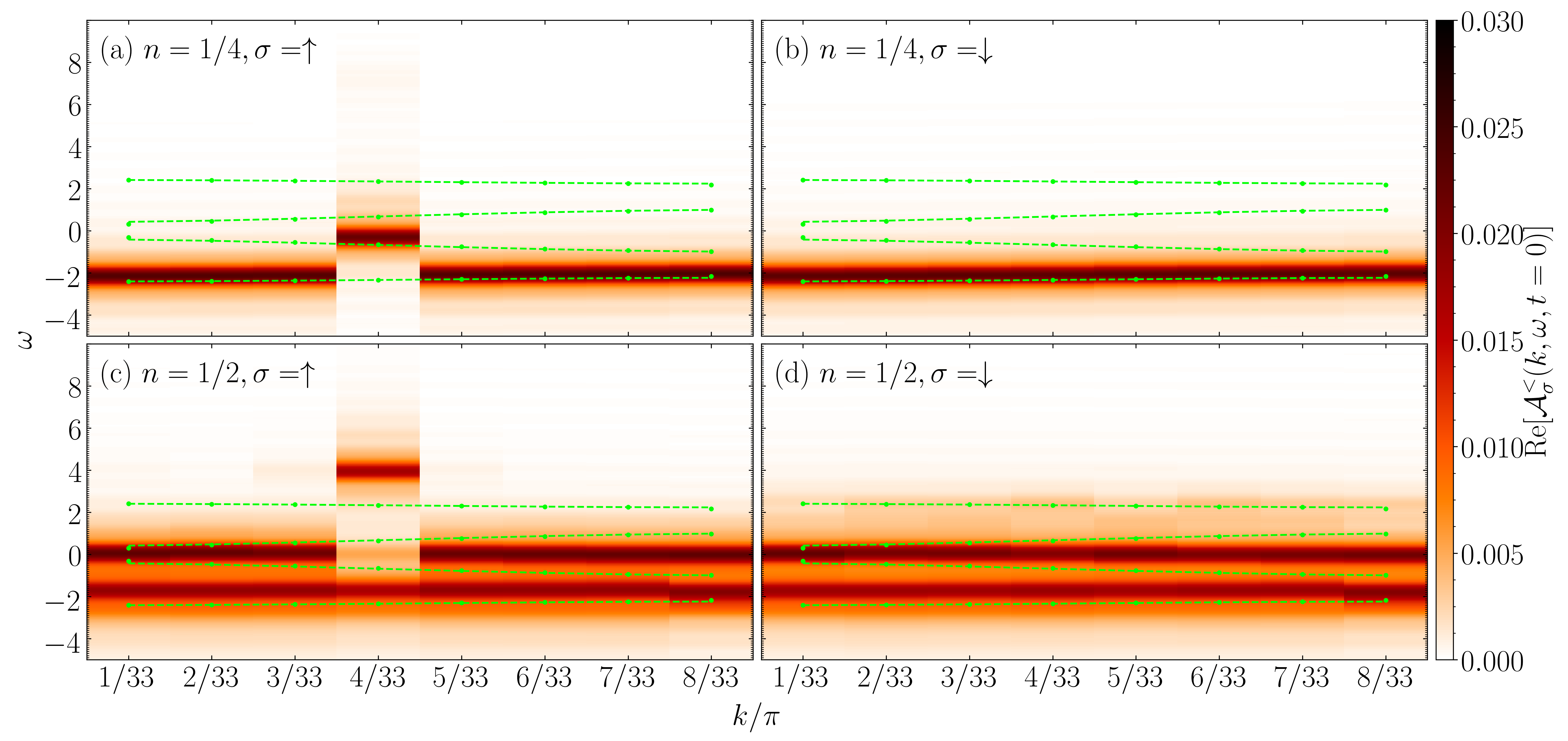}
	\caption{Single particle spectral functions $\mathcal{A}_{\sigma}^{<}\!\left(k,\omega,t=0\right)$  immediately after the electron-hole excitation \eqref{eq:lambda} for $\hat{H}^{4\Delta}$ with $U/t_{\mathrm{h}} = 4$, $L = 32, \Delta/t_{\mathrm{h}} = 2$ obtained with MPS for OBC  in the first BZ at quarter, (a) and (b) ($\chi_{\text{max}} = 500$), and half filling, (c) and (d), respectively. 
		(a) and (c) show the $\uparrow$-direction, (b) and (d) the $\downarrow$-direction. 
		The green dashed lines show the equilibrium band structure of the non-interacting system 
		calculated with PBC, the green dots correspond to the calculation with OBC. 
	}
	\label{fig:ex_t_0_cen_1_4_vs_1_2}
\end{figure*}
We consider an excitation, which resembles the absorption of a single photon with the energy corresponding to the gap at a fixed value of $k$. 
By construction, it induces a direct transition from the highest occupied band labeled by $\nu$ to the lowest unoccupied band $\nu+1$.  
This is possible here also for the interacting system, since the peak structure within the Hubbard bands of this correlated band insulator can be labeled by the $\nu$-values of the non-interacting system, as seen, e.g., in Fig.~\ref{fig:crosssection}.
However, care needs to be taken: i) due to the scattering between the electrons, the weight according to the quantum number $\nu$ is in general not strictly restricted to one $(k,\omega)$-point
; ii) in contrast to the non-interacting case, in which the complete weight of the electron is transferred to the higher band, correlation effects can cause some weight to remain in the lower Hubbard band.
Nevertheless, we find that for the correlated band insulators treated here, this is a useful modeling since the largest part of the weight is transferred to the higher band. 

We choose the excitation to affect only particles of the spin-$\uparrow$ direction.
More precisely, we take $k = 4\pi/33 \approx \pi/8$ for $\hat{H}^{4\Delta}$ and $k = 8\pi/33 \approx \pi/4$ for $\hat{H}^{2\Delta}$, since in this case finite size or boundary effects mainly affecting the edges of the first BZ (see Fig.~\ref{fig:eff_u}) are smallest and neglected in the following. 
The Green's functions according to Eq.~\eqref{eq:tsfs_g} are hence obtained by $\left|\Psi\right\rangle = \hat{\Lambda}\left|\text{GS}\right\rangle$ with
\begin{equation}
	\label{eq:lambda}
	\hat{\Lambda} = \hat{a}^{\dagger}_{\uparrow,(\nu+1),k}\hat{a}^{\phantom{\dagger}}_{\uparrow,\nu,k}, 
\end{equation}
with the operators $\hat{a}^{\left(\dagger\right)}_{\sigma,\nu,k}$ defined in Eq.~\eqref{eq:model_fold_back_4}.
For $\hat{H}^{4\Delta}$ we treat excitations with $\nu=2$ at half and $\nu=1$ at quarter filling, respectively. 
For $\hat{H}^{2\Delta}$ we consider only half filling and treat excitations with $\nu=1$.

\subsubsection*{Immediate effects}

\begin{figure*}
	\includegraphics[width=\textwidth]{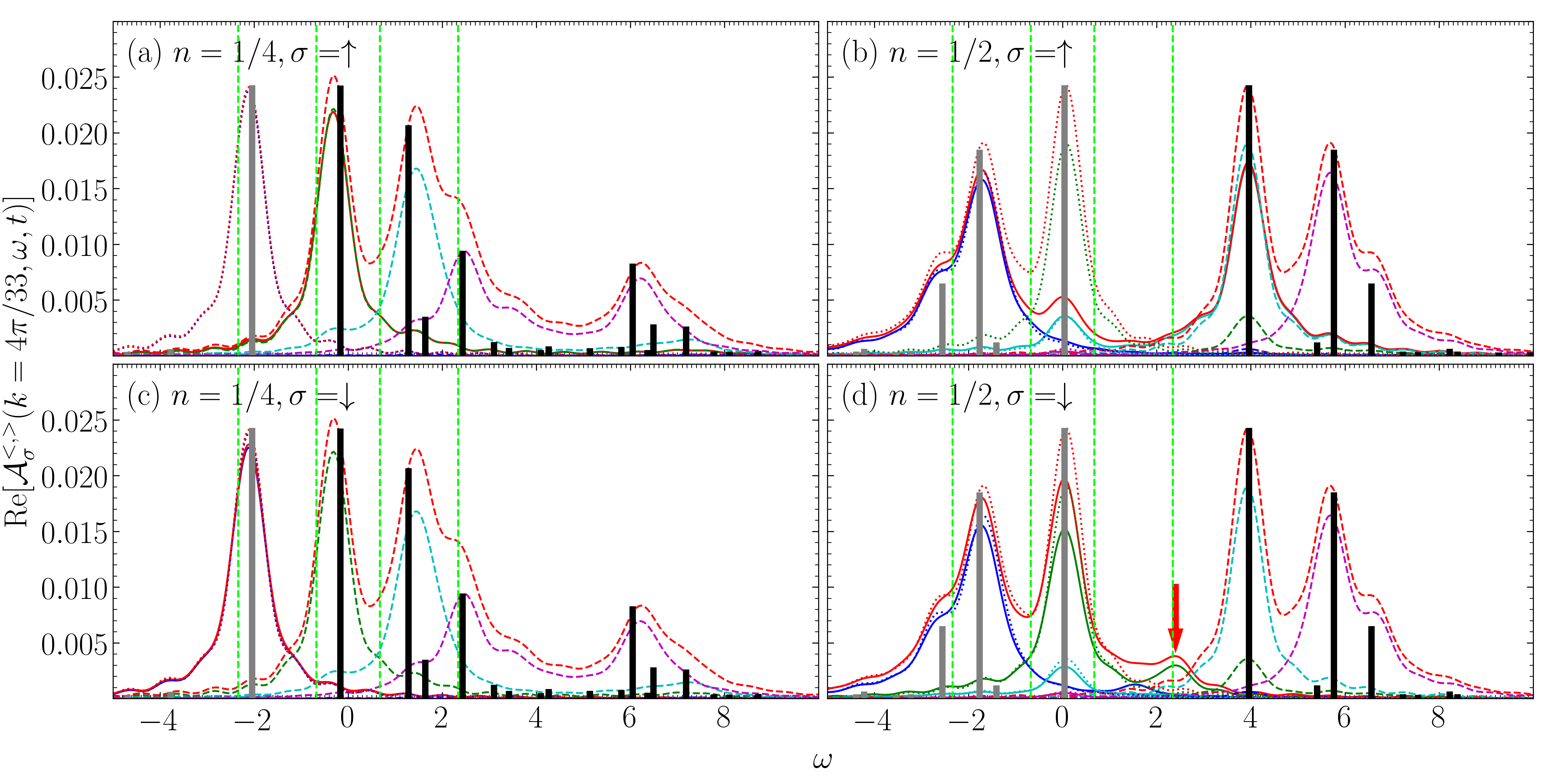}
	\caption{Cross section of the data shown in Fig.~\ref{fig:ex_t_0_cen_1_4_vs_1_2} at $k = 4\pi/33$ (solid lines) for quarter, (a) and (c) ($\chi_{\text{max}} = 500$) and half filling, (b) and (d). 
		The dotted (dashed) lines show the ground state results of Fig.~\ref{fig:crosssection} for the lesser (greater) spectral functions. 
		The total spectral function is shown in red, the contribution from the four band indices $\nu=1,2,3,4$ are given in blue, green, cyan, and magenta, respectively.  
		Vertical dashed light green lines: ground state peak positions in the non-interacting case obtained for PBC. 
		Vertical gray (black) bars: excitation energies for $\mathcal{A}^{<}_{\sigma}$ ($\mathcal{A}^{>}_{\sigma}$) in the ground state as obtained from Eq. \eqref{eq:lehmann_calc} for one unit cell, i.e. $L=4$. The heights of the bars correspond to their respective weights. All bars have been scaled such that in each cross section the largest bar takes a value of $0.8$ times the plot's maximum range in $y$-direction. 
		The red arrow highlights the in-gap spectral weight in $\mathcal{A}_{\downarrow}^{<}\!\left(k,\omega,t=0\right)$ (see text for further details).
	}
	\label{fig:ex_t_0_cen_1_4_vs_1_2_cs}
\end{figure*}
\begin{figure}
	\centering
	\includegraphics[width=0.48\textwidth]{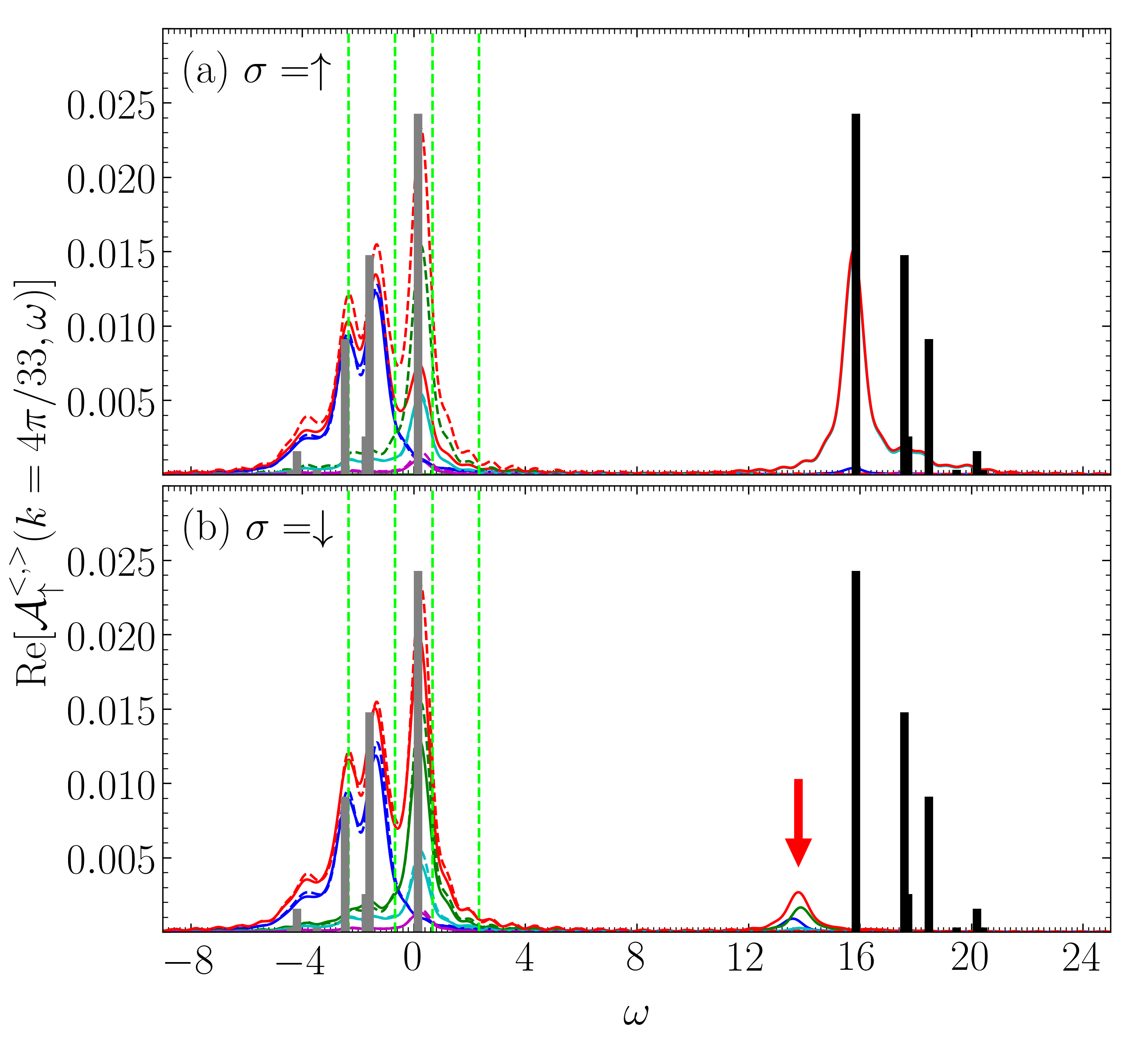}
	\caption{Cross section of $\mathcal{A}_{\sigma}^{<}\!\left(k,\omega,t\right)$ for $\hat{H}^{4\Delta}$ with $U/t_{\mathrm{h}} = 16$, $L = 32, \Delta/t_{\mathrm{h}} = 2$ at $k = 4\pi/33$ before the excitation ($t = - \infty$, dashed lines) and directly after the excitation ($t=0$, solid lines) for half filling. 
		The total spectral function is shown in red, the contribution from the four band indices $\nu=1,2,3,4$ are given in blue, green, cyan, and magenta, respectively.  
		Vertical dashed light green lines: ground state peak positions in the non-interacting case obtained for PBC. 
		Vertical gray (black) bars: excitation energies for $\mathcal{A}^{<}_{\sigma}$ ($\mathcal{A}^{>}_{\sigma}$) in the ground state as obtained from Eq. \eqref{eq:lehmann_calc} for one unit cell, i.e. $L=4$. The heights of the bars correspond to their respective weights. All bars have been scaled such that in each cross section the largest bar takes a value of $0.8$ times the plot's maximum range in $y$-direction. 
		The red arrow highlights the in-gap spectral weight in $\mathcal{A}_{\downarrow}^{<}\!\left(k,\omega,t=0\right)$. ($\chi_{\text{max}} = 500$.)
	}
	\label{fig:crossectionu16}
\end{figure}
\begin{figure}
	\begin{tabular}{l@{$\qquad$}r@{$\,$}c@{$\,$}l@{$\quad$}r@{$\,$}c@{$\,$}l}
		\multicolumn{7}{l}{(a) $\hat{H}^{4\Delta}$:} \\
		& $\left|\text{GS}\right>$ & $=$ & $\left|\downarrow\downarrow\uparrow\uparrow\right>$ & $E_{0}$ & $=$ & $-2\Delta$ \\
		\multicolumn{7}{l}{$\uparrow$-excitation:} \\
		& $\left|\Psi_{1}\right>$ & $=$ & $\left|\downarrow\updownarrow 0 \uparrow\right>$ & $E_{\Psi_{1}}$ & $=$ & $U-\Delta$ \\
		&$\hat{c}_{\downarrow}\left|\Psi_{1}\right>$ & $=$ & $\left|0\updownarrow 0\uparrow\right>$ & $E_{n=1}$ & $=$ & $U-\Delta/2$ \\
		& & & $\left|\downarrow\uparrow 0\uparrow\right>$ & $E_{n=2}$ & $=$ & $-\Delta/2$ \\
		\\
		\multicolumn{7}{l}{$\Rightarrow \quad E_{\Psi_{1}}-E_{n} = \begin{cases}-\Delta/2, & n=1\\ U-\Delta/2, & n=2\end{cases}$} \\
		\\
		& $\left|\Psi_{2}\right>$ & $=$ & $\left|\updownarrow\updownarrow 0 0\right>$ & $E_{\Psi_{2}}$ & $=$ & $2U$ \\
		&$\hat{c}_{\downarrow}\left|\Psi_{2}\right>$ & $=$ & $\left|\uparrow\updownarrow 00\right>$ & $E_{n=1}$ & $=$ & $U+\Delta/2$ \\
		& & & $\left|\updownarrow\uparrow 00\right>$ & $E_{n=2}$ & $=$ & $U+\Delta/2$ \\
		\\
		\multicolumn{7}{l}{$\Rightarrow \quad E_{\Psi_{2}}-E_{n=1,2} = U - \Delta / 2$} \\
		\\
		\multicolumn{7}{l}{(b) $\hat{H}^{2\Delta}$:} \\
		& $\left|\text{GS}\right>$ & $=$ & $\left|\downarrow\uparrow\right>$ & $E_{0}$ & $=$ & $-\Delta$ \\
		\multicolumn{7}{l}{$\uparrow$-excitation:} \\
		& $\left|\Psi\right>$ & $=$ & $\left|\updownarrow 0\right>$ & $E_{\Psi}$ & $=$ & $U$ \\
		&$\hat{c}_{\downarrow}\left|\Psi\right>$ & $=$ & $\left|\uparrow 0\right>$ & $E_{n=1}$ & $=$ & $+\Delta/2$ \\
		\\
		\multicolumn{7}{l}{$\Rightarrow \quad E_{\Psi}-E_{n=1} = U - \Delta / 2$} \\
	\end{tabular}
	\caption{Sketch of the states and energies needed to compute $\mathcal{A}_{\downarrow}^{<}\!\left(\omega\right)$ after an excitation in the $\uparrow$-direction, which shifts the particle by one site, in the atomic limit $\hopping=0$ (a) for $\hat{H}^{4\Delta}$ and (b) for $\hat{H}^{2\Delta}$. 
		The difference in energies corresponds to the position of the mid-gap peak, here for both cases at $U-\Delta/2$. 
	}
	\label{fig:sketch}
\end{figure}
\begin{figure*}
	\includegraphics[width=\textwidth]{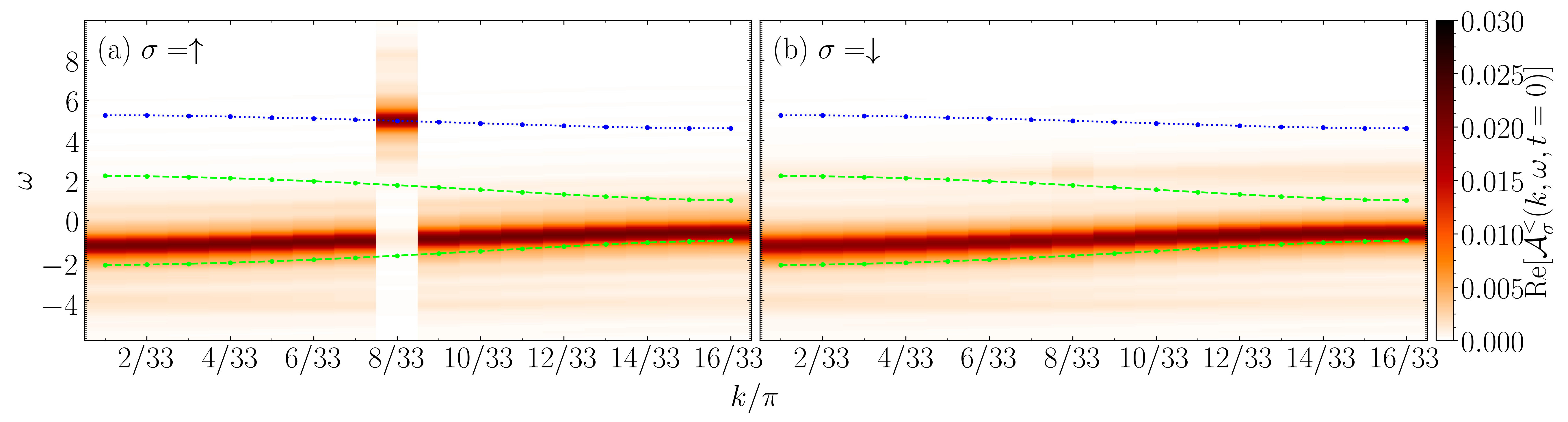}
	\caption{Single particle spectral functions $\mathcal{A}_{\sigma}^{<}\!\left(k,\omega,t=0\right)$  immediately after the electron-hole excitation \eqref{eq:lambda} for $\hat{H}^{2\Delta}$ with $U/t_{\mathrm{h}} = 4$, $L = 32, \Delta/t_{\mathrm{h}} = 2$ obtained with MPS for OBC  in the first BZ at half filling. 
		(a) shows the $\uparrow$-direction, (b) the $\downarrow$-direction. 
		The green dashed lines show the equilibrium band structure of the non-interacting system 
		calculated with PBC, the green dots correspond to the calculation with OBC. 
		The blue dotted line depicts the maxima of $\mathcal{A}_{\sigma}^{>}\!\left(k,\omega\right)$ extracted from Fig.~\ref{fig:eff_u_2pcmo}(b).
	}
	\label{fig:m_ex_t_0_u_cen_2pcmo}
\end{figure*}
\begin{figure}[b!]
	\includegraphics[width=0.48\textwidth]{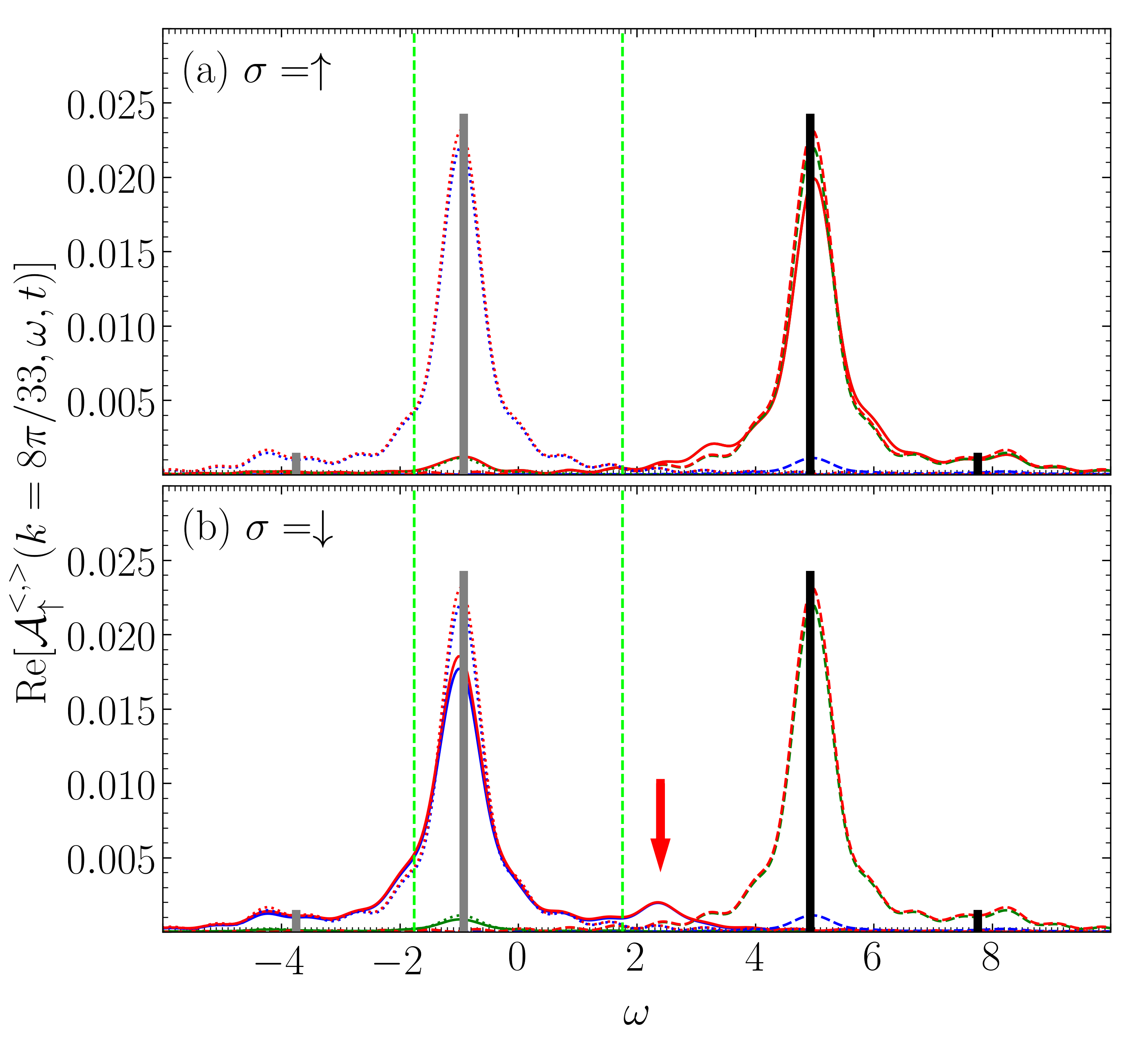}
	\caption{Cross section of the data shown in Fig.~\ref{fig:m_ex_t_0_u_cen_2pcmo} at $k = 8\pi/33$ (solid lines) for half filling. 
		The dotted (dashed) lines show the ground state results of Fig.~\ref{fig:crosssection_2pcmo} for the lesser (greater) spectral functions. 
		The total spectral function is shown in red, the contribution from the two band indices $\nu=1,2$ are given in blue and green, respectively.  
		Vertical dashed light green lines: ground state peak positions in the non-interacting case obtained for PBC. 
		Vertical gray (black) bars: excitation energies for $\mathcal{A}^{<}_{\sigma}$ ($\mathcal{A}^{>}_{\sigma}$) in the ground state as obtained from Eq. \eqref{eq:lehmann_calc} for one unit cell, i.e. $L=2$. The heights of the bars correspond to their respective weights. All bars have been scaled such that in each cross section the largest bar takes a value of $0.8$ times the plot's maximum range in $y$-direction. 
		The red arrow highlights the in-gap spectral weight in $\mathcal{A}_{\downarrow}^{<}\!\left(k,\omega,t=0\right)$.
	}
	\label{fig:m_ex_t_0_u_cen_2pcmo_cs}
\end{figure}
First, we investigate the effect directly after the excitation, i.e., at $t=0$.
In Fig.~\ref{fig:ex_t_0_cen_1_4_vs_1_2} we show the results for the spectral function for $\hat{H}^{4\Delta}$ at $U/t_{\mathrm{h}} = 4$ and $\Delta/t_{\mathrm{h}} = 2$ for both, half and quarter filling. 
In the $\uparrow$-direction the weight in the upper occupied band at $k = 4\pi/33$ is significantly reduced, as desired, and at the same $k$-value states at higher energies get populated, with a clear maximum in the lowest `band' within the upper Hubbard band.

Note that for $U=0$ -- up to a small weight due to the limited resolution -- a perfect transfer of weight is obtained from the $\nu$- to the $\nu+1$-band and that the spectral function after the excitation is found numerically to be constant in time, as expected. 
However, this is only the case when applying the $k$-space transform to the eigenstates of the non-interacting Hamiltonian, see App.~\ref{app:obc}. 
This is not the case when performing the transform to other seemingly suitable bases, e.g. for OBC to a plane-wave basis or a simple $\sin$-transform \cite{benthien_sin}. 
Even though in equilibrium this typically leads to (small) finite size corrections, when dealing with the time-dependent non-equilibrium spectral function it is important to work in the correct eigenbases in order to avoid artificial time dependencies.

At $U=0$, the spin-$\downarrow$ direction is completely unaffected, due to the absence of interactions between both spin channels. 
From Figs.~\ref{fig:ex_t_0_cen_1_4_vs_1_2}(a) and (b) we find that -- apart from the slight renormalization discussed before -- at quarter filling the situation in excellent approximation resembles the non-interacting case.
This further indicates that at this filling and value of the parameters  interaction effects are not dominant. 
However at half filling, shown in Figs.~\ref{fig:ex_t_0_cen_1_4_vs_1_2}(c) and (d), the behaviour differs significantly from the non-interacting case:
a small, but finite weight remains in $\mathcal{A}_{\uparrow}^{<}\!\left(k,\omega,t=0\right)$, and the population in $\mathcal{A}_{\uparrow}^{>}\!\left(k,\omega,t=0\right)$ is smeared out to higher energies and also weakly to neighboring $k$-values, both of which we associate to the present scattering between the electrons.
Most prominently, however, we find the electrons in the $\downarrow$-direction to be affected as well, even though we did not apply an excitation there. 
We observe a new band in between the lower and the upper Hubbard band, at an energy $\omega \approx 2.5$.
Note that at this value of $\omega$ there is no weight in the FD- or MPS-treatment of the equilibrium spectral function. 

We further illustrate these findings at half filling on the corresponding cross sections at $k = 4\pi/33$ in Fig.~\ref{fig:ex_t_0_cen_1_4_vs_1_2_cs}.
The dominant effect is that the spectral weight of the $\nu=2$ contribution to $\mathcal{A}_{\uparrow}^{<}\!\left(k,\omega\right)$, which makes for most of the corresponding band, has vanished after application of the operator~\eqref{eq:lambda}.
We find that it has been transferred mainly to the $\nu=3$ contribution of $\mathcal{A}_{\uparrow}^{>}\!\left(k,\omega\right)$, and both lines almost perfectly overlap.
The remnant of weight in the highest occupied band of  $\mathcal{A}_{\uparrow}^{<}\!\left(k,\omega\right)$ is seen to mostly come from its $\nu=3$ contribution, which is unaffected in our modeled excitation \eqref{eq:lambda}. 
Since this contribution is zero in the non-interacting case, this feature can be associated to the stronger interaction effects at half filling, which were essentially absent at quarter filling. 
Turning our attention to $\mathcal{A}_{\downarrow}^{<}\!\left(k,\omega,t=0\right)$ in Fig.~\ref{fig:ex_t_0_cen_1_4_vs_1_2_cs}(d) we find that also in this case the $\nu=2$ contribution of $\mathcal{A}_{\downarrow}^{<}\!\left(k,\omega\right)$ has dropped quite significantly and the new feature appears as indicated by the red arrow in Fig.~\ref{fig:ex_t_0_cen_1_4_vs_1_2_cs}(d) at the mid-gap energy value $\omega \approx 2.5$. 

This is a remarkable finding, which we further analyze by changing the value of $U$.
In Fig.~\ref{fig:crossectionu16}, we show the same cross sections as in Fig.~\ref{fig:ex_t_0_cen_1_4_vs_1_2_cs} at half filling for $U=16$.
Here, the new midgap-feature in the spin-$\downarrow$ direction is clearly visible.
Its energy is at $\omega \approx U-\Delta$, which appears to be outside the upper Hubbard band. 
Such a feature can be obtained in the atomic limit $\hopping \to 0$, c.f. Fig.~\ref{fig:sketch}: 
computing $\mathcal{A}^{<}\!\left(k,\omega\right)$ after the electron-hole excitation for a single unit cell, one finds a signal in the spin-$\downarrow$-direction at $\omega=U-\Delta/2$, which lies in the gap, and one in the spin-$\uparrow$-direction at $\omega=U+\Delta/2$, which can lie inside the upper Hubbard band.
In the results of Figs.~\ref{fig:ex_t_0_cen_1_4_vs_1_2} and \ref{fig:m_ex_t_0_u_cen_2pcmo}, which are away from the atomic limit, the aforementioned signal in the $\downarrow$-direction is obtained, but it is more difficult or impossible to identify one in the $\uparrow$-direction, even when changing the value of $\Delta/\hopping$, see Fig.~\ref{fig:a_oc_gs_vs_t_0_4_delta_8} for results at $\Delta/\hopping=8$. 

\begin{figure}[t!]
	\includegraphics[width=0.48\textwidth]{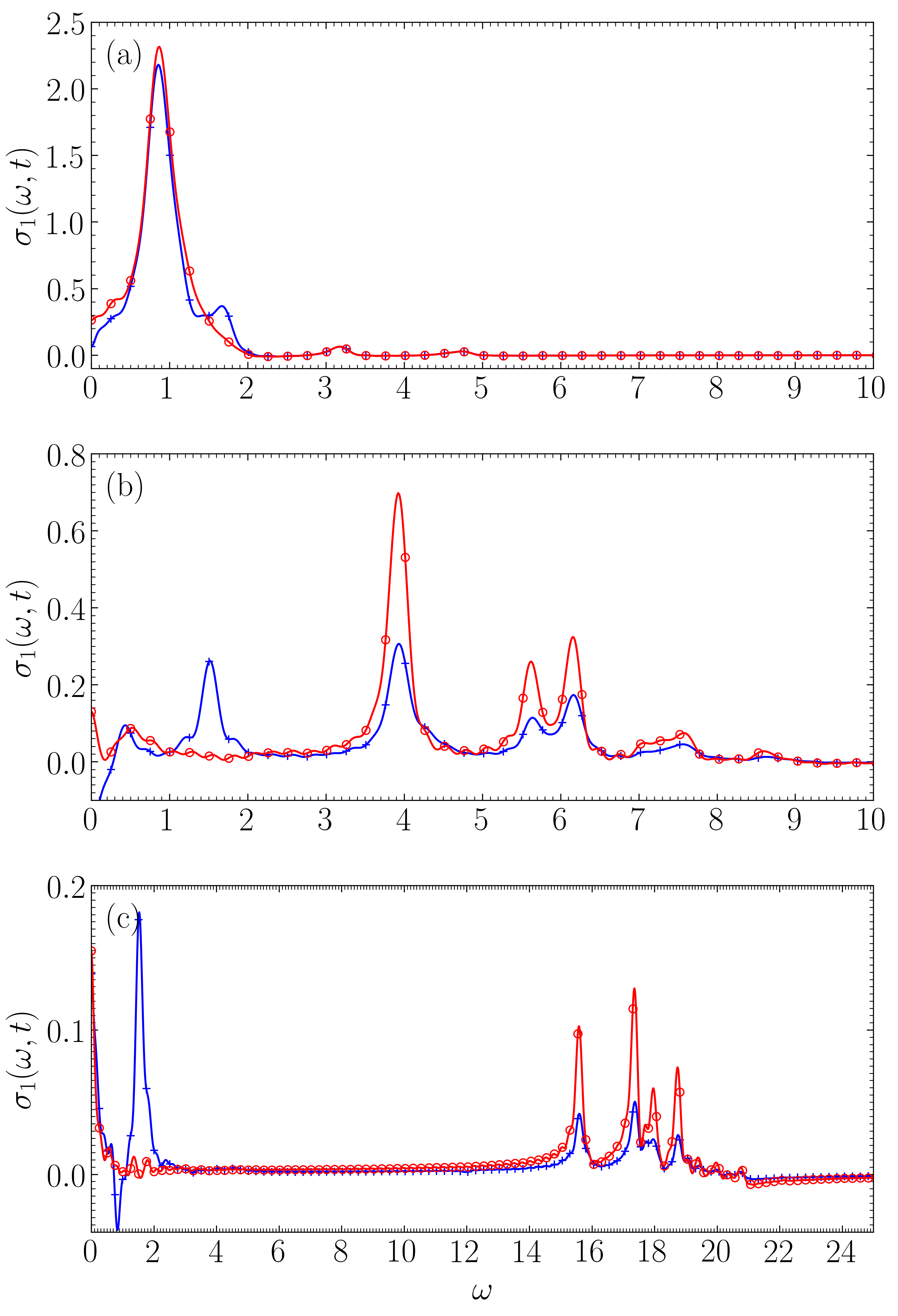} 
	\caption{Real part of the optical conductivity $\sigma_{1}\!\left(\omega,t\right)$ for $\hat{H}^{4\Delta}$ with $\Delta/\hopping=2$ and $U/\hopping=0, 4,$ and $16$ ((a) to (c)) at half filling in the ground state, i.e. $t=-\infty$, (red) and immediately after the excitation, i.e. $t=0$ (blue). 
		Data points: Obtained from Fourier transforming according to \eqref{eq:methods_oc_ft_a} and \eqref{eq:methods_oc_ft_j}, respectively, with damping $\tilde{\eta}=0.1$. Solid lines: Application of $16$ times zero padding.}
	\label{fig:oc_gs_vs_t_0_4}
\end{figure}
\begin{figure}[t!]
	\includegraphics[width=0.48\textwidth]{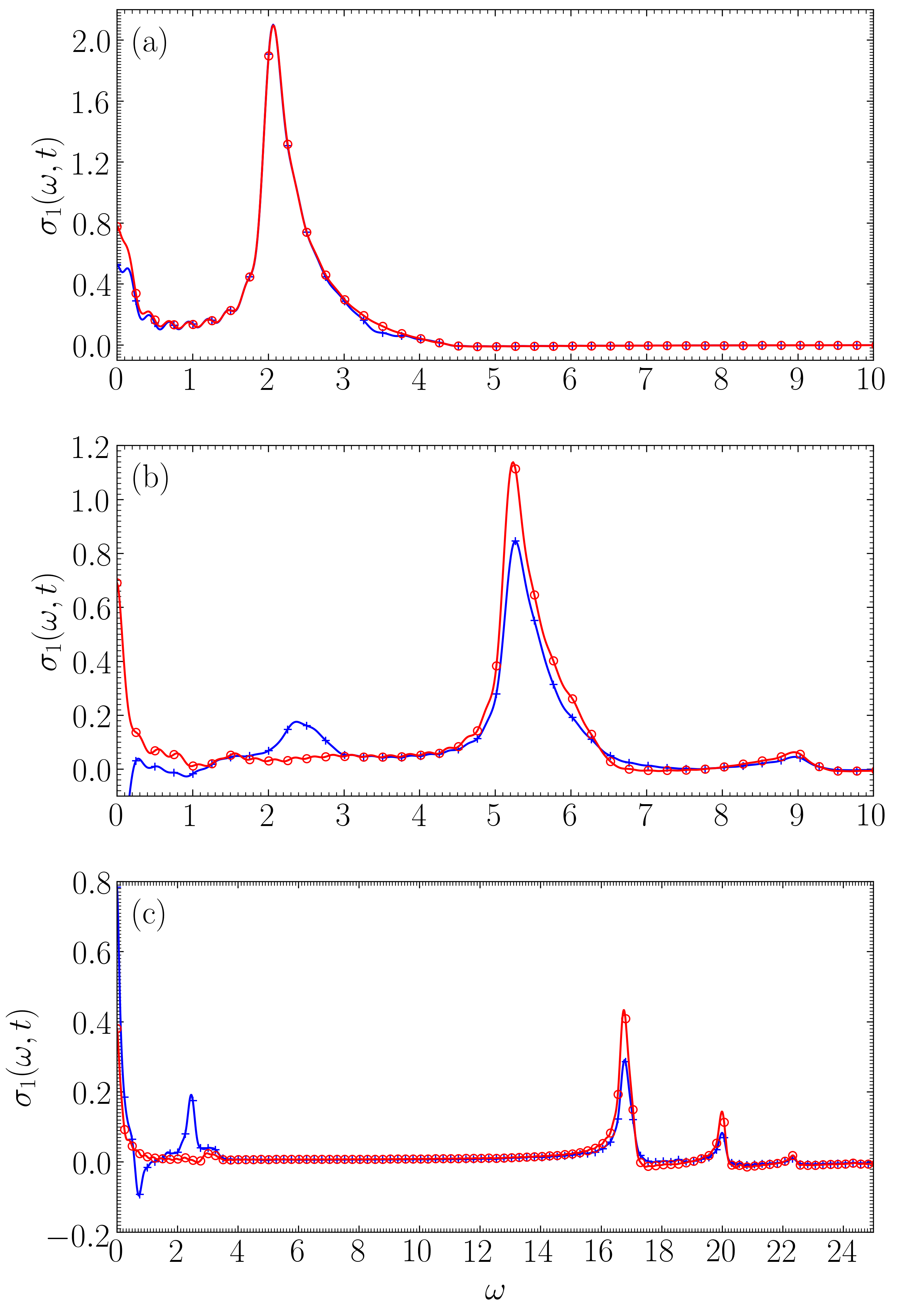} 
	\caption{Real part of the optical conductivity $\sigma_{1}\!\left(\omega,t\right)$ for $\hat{H}^{2\Delta}$ with $\Delta/\hopping=2$ and $U/\hopping=0, 4,$ and $16$ ((a) to (c)) at half filling in the ground state, i.e. $t=-\infty$, (red) and immediately after the excitation, i.e. $t=0$ (blue). 
		Data points: Obtained from Fourier transforming according to \eqref{eq:methods_oc_ft_a} and \eqref{eq:methods_oc_ft_j}, respectively, with damping $\tilde{\eta}=0.1$. Solid lines: Application of $16$ times zero padding.}
	\label{fig:oc_gs_vs_t_0_2}
\end{figure}

We present our results for $\hat{H}^{2\Delta}$ for which we show the full lesser spectral functions after the excitation $\mathcal{A}^{<}\!\left(k,\omega,t=0\right)$ in Fig.~\ref{fig:m_ex_t_0_u_cen_2pcmo}. 
Due to the simpler band structure, the effects of the excitation in $\hat{H}^{2 \Delta}$ are easier to analyze.
First, we note that in the $\uparrow$-direction in Fig.~\ref{fig:m_ex_t_0_u_cen_2pcmo}(a) the remaining weight in the lower band is significantly less than for the situation at half filling in $\hat{H}^{4\Delta}$, c.f. Fig.~\ref{fig:ex_t_0_cen_1_4_vs_1_2}(c).
Second, in addition we find a mid-gap band at $\omega \approx 2.4$ in the $\downarrow$-direction, as seen in Fig.~\ref{fig:m_ex_t_0_u_cen_2pcmo}(b), the only difference being that in this case the weight in the mid-gap band is not as equally distributed as in Fig.~\ref{fig:ex_t_0_cen_1_4_vs_1_2}(b) but rather more accumulated at larger momenta. 
This additional feature is also seen in the cross section at $k = 8\pi/33$ in Fig.~\ref{fig:m_ex_t_0_u_cen_2pcmo_cs}. 
As expected, in the $\uparrow$-direction the dominant $\nu = 1$ contribution $\mathcal{A}_{\uparrow}^{<}\!\left(k,\omega\right)$ has vanished entirely in $\mathcal{A}_{\uparrow}^{<}\!\left(k,\omega,t=0\right)$ and is transfered to the energy corresponding to the $\nu = 2$ contribution of $\mathcal{A}_{\uparrow}^{>}\!\left(k,\omega\right)$, c.f. Fig.~\ref{fig:m_ex_t_0_u_cen_2pcmo_cs}(a). 
Due to the simpler band structure, there is only a slight remnant caused by the $\nu = 2$ contribution to $\mathcal{A}_{\uparrow}^{<}\!\left(k,\omega\right)$. 
As for the $\downarrow$-direction in Fig.~\ref{fig:m_ex_t_0_u_cen_2pcmo_cs}(b), the mid-gap state does not correspond to any frequency of $\mathcal{A}_{\downarrow}^{>}\!\left(k,\omega\right)$ and is even further away from its $\nu = 2$ contribution as in the case of $\hat{H}^{4\Delta}$, c.f. Fig.~\ref{fig:ex_t_0_cen_1_4_vs_1_2_cs}(d). 
We attribute the other deviations between the spectral functions before and after the excitation again to scattering processes.
A midgap state is again found in the atomic limit, see Fig.~\ref{fig:sketch}. 
In summary, these findings indicate that the formation of mid-gap bands is generically obtained in photo excited Hubbard systems with a magnetic superstructure.

\subsection{Optical conductivity and exciton states}
\label{sec:results_oc}

\begin{figure*}
	\includegraphics[width=\textwidth]{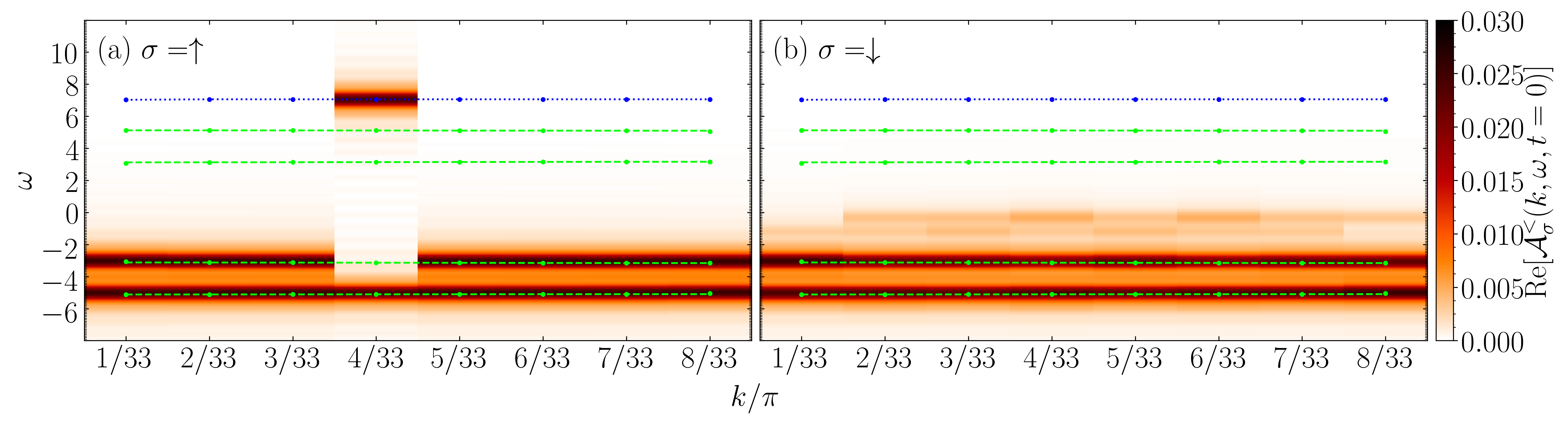}
	\caption{Single particle spectral functions $\mathcal{A}_{\sigma}^{<}\!\left(k,\omega,t=0\right)$  immediately after the electron-hole excitation \eqref{eq:lambda} for $\hat{H}^{4\Delta}$ with $U/t_{\mathrm{h}} = 4$, $L = 32, \Delta/t_{\mathrm{h}} = 8$ obtained with MPS for OBC  in the first BZ at half filling. 
		(a) shows the $\uparrow$-direction, (b) the $\downarrow$-direction. 
		The green dashed lines show the equilibrium band structure of the non-interacting system 
		calculated with PBC, the green dots correspond to the calculation with OBC. 
		The blue dotted line depicts the maxima of $\mathcal{A}_{\sigma}^{>}\!\left(k,\omega\right)$. ($\chi_{\text{max}} = 500$.)}
	\label{fig:a_oc_gs_vs_t_0_4_delta_8}
\end{figure*}
\begin{figure}
	\includegraphics[width=0.48\textwidth]{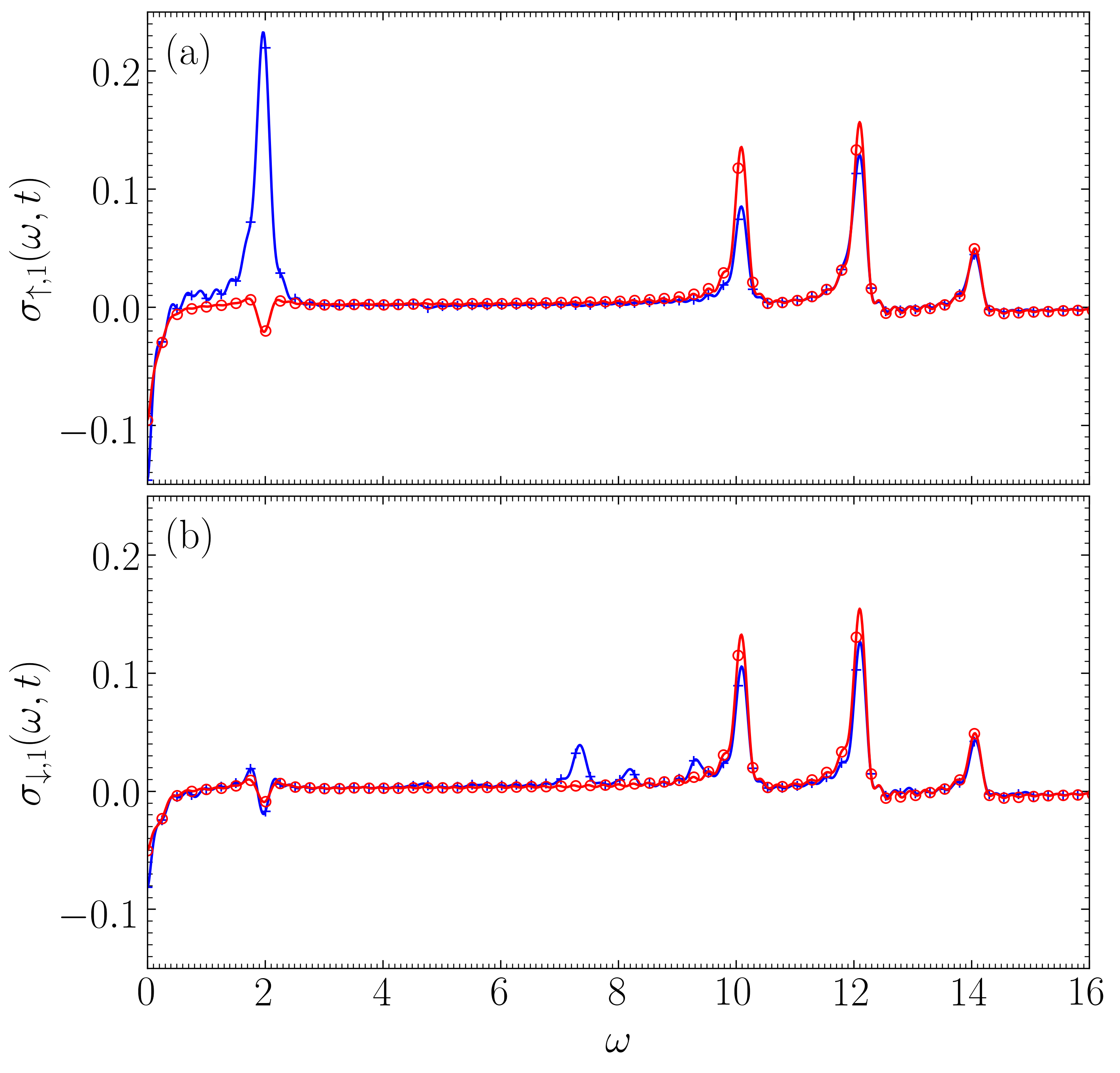}
	\caption{Real part of the spin-resolved optical conductivity $\sigma_{1}\!\left(\omega,t\right)$ for $\hat{H}^{4\Delta}$ with $\Delta/\hopping=8$ and $U/\hopping=4$ at half filling. (a): $\uparrow$-direction; (b): $\downarrow$-direction. Red denotes the ground state, i.e. $t=-\infty$, blue the results immediately after the excitation, i.e. $t=0$.  
		Data points: Obtained from Fourier transforming according to \eqref{eq:methods_oc_ft_a} and \eqref{eq:methods_oc_ft_j}, respectively, with damping $\tilde{\eta}=0.1$. Solid lines: Application of $16$ times zero padding.}
	\label{fig:oc_gs_vs_t_0_4_delta_8}
\end{figure}

We further analyze the situation by computing the optical conductivity at half filling before and after the excitation, as shown for $\hat{H}^{4\Delta}$ in Fig.~\ref{fig:oc_gs_vs_t_0_4} at values of  $\Delta/\hopping = 2$, $U/\hopping = 0, \, 4$ and $16$, and in Fig.~\ref{fig:oc_gs_vs_t_0_2} for the same values of $\Delta/\hopping$ and $U/\hopping$ for $\hat{H}^{2\Delta}$. 
In addition we present in Figs.~\ref{fig:a_oc_gs_vs_t_0_4_delta_8} and~\ref{fig:oc_gs_vs_t_0_4_delta_8} reference results of the spectral function and the optical conductivity after the excitation for $\hat{H}^{4\Delta}$ at  $\Delta/\hopping=8$.
Note that for the optical conductivity we will not discuss features at frequencies $\omega \to 0$,  
since our approach has the largest uncertainties there, as discussed in Sec.~\ref{sec:methods_oc}.

Before the excitation, we find in both models multiple peaks at energies $\omega \sim U/\hopping$ or higher, whose values for the most prominent peaks are listed in Tab.~\ref{tab:oc_gs_4_2}.
\begin{table}
	\caption{Peak positions in the optical conductivity $\omega_{\sigma_{1}}$ vs. peak-to-peak distance in the spectral function $\omega_{\mathcal{A}}$ for $\hat{H}^{4\Delta}$ and $\hat{H}^{2\Delta}$  at $\Delta=2$ and various $U$. Note that at $U=0$ the dispersion in the spectral function is strongest so that the results may deviate.}
	
	\begin{tabular}{c@{$\quad$}c@{$\quad$}c@{$\qquad$}c@{$\quad$}c@{$\quad$}}
		& \multicolumn{2}{c}{$\hat{H}^{4\Delta}$} & \multicolumn{2}{c}{$\hat{H}^{2\Delta}$} \\
		\toprule
		$U$ & $\omega_{\sigma_{1}}$ & $\omega_{\mathcal{A}}$ & $\omega_{\sigma_{1}}$ & $\omega_{\mathcal{A}}$  \\
		\midrule
		0&0.87 & 0.62 & 2.07 & 2.01\\
		&3.17 & 2.71 &&\\
		&4.78 & 4.81 &&\\
		\midrule
		4&3.93&3.93 & 5.23 & 5.2 \\
		&5.62&5.76 & 8.95 & 8.64 \\
		&6.16&6.51 &&\\
		&8.60&8.33 &&\\
		\midrule
		16&15.60&15.18 & 16.75 & 16.72\\
		&17.35&16.76 & 20.0 & 19.94\\
		&17.95&17.73 & 22.3 & 21.5\\
		&18.73&19.30 &&\\
		\bottomrule
	\end{tabular}
	\label{tab:oc_gs_4_2}
\end{table}
These values correspond rather well to the energy differences between the peak positions of the highest occupied band in $\mathcal{A}^<(k,\omega)$ and the peak positions of the empty bands in $\mathcal{A}^>(k,\omega)$. 
Hence, the peaks seen in the equilibrium optical conductivity correspond to transferring a particle from the lower Hubbard band to the higher Hubbard band. 
This is obtained for $\hat{H}^{2\Delta}$ and for $\hat{H}^{4\Delta}$.

However, in all cases when $\Delta>0$, additional peaks appear immediately after the excitation. 
At equilibrium, peaks in the optical conductivity below the Mott gap indicate the formation of excitons in interacting electron systems~\cite{Excitons2001}. 
Here, the question arises if this is true also after an excitation; in case of exciton formation, we expect additional signals in both the optical conductivity and the spectral function.
The latter is expected to show a feature at an energy, which corresponds to the conduction band (upper Hubbard band) minus the binding energy $E_b$ of the exciton\cite{christiansen2019,perfetto_melting_noneq_exc}.
In a Mott insulator, the relation between $E_{b}$, the Mott gap $E_{M}$ and the peak position of the exciton signal in the optical conductivity $\omega_{\text{exc}}$ is~\cite{Excitons2001,Jeckelmann2003,Benthien2005}  $E_{b}=E_{M}-\omega_{\text{exc}}$.
The Mott gap can be read off directly from the spectral function, so that we do not need further computations and we can directly check for this expectation.  

We find that we need to differentiate between $\hat{H}^{4\Delta}$ and $\hat{H}^{2\Delta}$. 
For $\hat{H}^{4\Delta}$ we find an additional peak at $\omega \approx 2$ in the optical conductivity independent of the interaction strength $U$, see Fig.~\ref{fig:oc_gs_vs_t_0_4}. 
Since at $U=0$ the spin-$\uparrow$ and spin-$\downarrow$ electrons are not coupled, there is no additional signal in the spectral function.
Thus, in general its origin cannot be traced back to the formation of excitons. 
Hence, additional peaks in the optical conductivity after a photo excitation do not necessarily indicate the formation of excitons. 

This is further supported by the findings presented in Fig.~\ref{fig:oc_gs_vs_t_0_4_delta_8} where we show the optical conductivity before and after the excitation for $U=4$ and $\Delta=8$ and resolved for both spin directions. 
We notice that apart from the peak at $\omega \approx 2$, here we encounter other new peaks after the excitation at $\omega \approx 7.5, 8, 9.5$. 
While the peak at $\omega \approx 2$ is only in the $\uparrow$-direction, the peaks at $\omega \approx 7.5, 8, 9.5$ are only in the $\downarrow$-direction. 
These correspond to additional features in $\mathcal{A}^{<}_{\downarrow}\!\left(k,\omega,t>0\right)$, and hence can be related to the formation of three $\downarrow$-excitons.
However, they do not show at the expected energies $E_{M}-E_{b}$ above the lower Hubbard band but rather at $E_{M}-\omega_{\text{exc}}$.

For finite $U$ and $\Delta=2$ the mid-gap states are at energies $\omega \approx 2$ below the upper Hubbard band, c.f. Figs.~\ref{fig:ex_t_0_cen_1_4_vs_1_2_cs}(d) and \ref{fig:crossectionu16}(b). 
Figure~\ref{fig:oc_gs_vs_t_0_4} can then be interpreted such that two effects are superimposed and the additional peak at $\omega\approx 2$ is caused by both, the exciton and the particular band structure of $\hat{H}^{4\Delta}$:
The upper and lower Hubbard bands show two internal bands, which according to Eq.~\eqref{eq:model_disp_4} are separated by $\omega \approx 2$ for large $\Delta$. 
When an $\uparrow$-electron is in the third band, it takes $\omega\approx 2$ to shift it to the fourth band, explaining the peak in Fig.~\ref{fig:oc_gs_vs_t_0_4} at $U=0$. 
A more detailed analysis shows this peak to only stem from the spin-$\uparrow$ direction. 
Generically, in the non-interacting case, we find no such peak in the spin-$\downarrow$ direction, and also at finite $U$ the major contribution is only in the spin-$\uparrow$ direction, see Fig.~\ref{fig:oc_gs_vs_t_0_4_delta_8} as an illustrative example. 

For $\hat{H}^{2\Delta}$ the situation is clearer due to the simpler band structure.
We again find peaks at $\omega \approx 2$ after the excitation, but this time it depends on $U$, c.f. Fig.~\ref{fig:oc_gs_vs_t_0_2}. 
At $U>0$, these additional peaks are again only in the $\downarrow$-direction.
However, note that for $U=0$ also the ground state optical conductivity shows a peak at $\omega \approx 2$ in contrast to the cases of finite $U$, further indicating they cannot be of the same origin. 
As for $\hat{H}^{4\Delta}$ we find the exciton band at energies approximately $E_{M}-\omega_{\text{exc}}$ above the lower Hubbard band in the spectral function for the $\downarrow$-direction. 
Note however that this new band is nearly dispersionless, or at least it has a smaller curvature than the upper Hubbard band. 

\begin{figure}
	\begin{center}
		\flushleft{(a) excitation}
		\ifthenelse{\boolean{buildtikzpics}}
		{
			\tikzsetnextfilename{img/tikz_ex}
			\begin{tikzpicture}
			\begin{scope} [node distance = 0.1 and 0.1]
			\node (g2) { \begin{tikzpicture} \bspGhost{0} \end{tikzpicture} };
			\node (g3) [right = of g2] { \begin{tikzpicture} \bspGhost{0} \end{tikzpicture} };
			\node (g4) [right = of g3] { \begin{tikzpicture} \bspGhost{0} \end{tikzpicture} };
			\node (g5) [right = of g4] { \begin{tikzpicture} \bspGhost{0} \end{tikzpicture} };
			\node (g6) [right = of g5] { \begin{tikzpicture} \bspGhost{0} \end{tikzpicture} };
			\node (g7) [right = of g6] { \begin{tikzpicture} \bspGhost{0} \end{tikzpicture} };
			\node (g8) [right = of g7] { \begin{tikzpicture} \bspGhost{0} \end{tikzpicture} };
			\node (s2) [below = of g2] { \begin{tikzpicture} \bspUp{0} \end{tikzpicture} };
			\node (s4) [below = of g4, draw=red, circle, yshift=0.3em] { \begin{tikzpicture}[baseline, yshift=0.0em] \bspNone{0} \end{tikzpicture} };
			\node (s6) [below = of g6] { \begin{tikzpicture} \bspUp{0} \end{tikzpicture} };
			\node (s3) [above = of g3] { \begin{tikzpicture} \bspDown{0} \end{tikzpicture} };
			\node (s5) [above = of g5] { \begin{tikzpicture} \color{red} \bspUpDown{0} [baseline, yshift=-0.65em] \color{black} \bspNone{0} \end{tikzpicture} };
			\node (s7) [above = of g7] { \begin{tikzpicture} \bspDown{0} \end{tikzpicture} };
			\node (t3) [above = of s3] { $-\Delta/2$ };
			\node (t5) [above = of s5, yshift=-0.4em] { $U$ };
			\node (t7) [above = of s7] { $-\Delta/2$ };
			\node (t2) [below = of s2] { $-\Delta/2$ };
			\node (t4) [below = of s4, yshift=0.3em] { $0$ };
			\node (t6) [below = of s6] { $-\Delta/2$ };
			\node (t7) [right = of s6, draw=black] { $E=U-2\Delta$ };
		\end{scope}
	\end{tikzpicture} \\[2em]
}
{
	\includegraphics{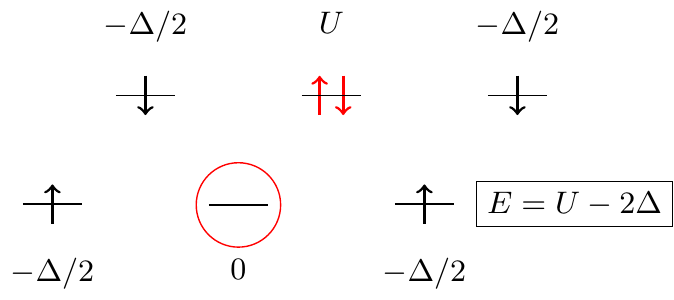}
}
\flushleft{(b) hole hopping}
\ifthenelse{\boolean{buildtikzpics}}
{
	\tikzsetnextfilename{img/tikz_hop_hole}
	\begin{tikzpicture}
	\begin{scope} [node distance = 0.1 and 0.1]
	\node (g2) { \begin{tikzpicture} \bspGhost{0} \end{tikzpicture} };
	\node (g3) [right = of g2] { \begin{tikzpicture} \bspGhost{0} \end{tikzpicture} };
	\node (g4) [right = of g3] { \begin{tikzpicture} \bspGhost{0} \end{tikzpicture} };
	\node (g5) [right = of g4] { \begin{tikzpicture} \bspGhost{0} \end{tikzpicture} };
	\node (g6) [right = of g5] { \begin{tikzpicture} \bspGhost{0} \end{tikzpicture} };
	\node (g7) [right = of g6] { \begin{tikzpicture} \bspGhost{0} \end{tikzpicture} };
	\node (g8) [right = of g7] { \begin{tikzpicture} \bspGhost{0} \end{tikzpicture} };
	\node (s2) [below = of g2] { \begin{tikzpicture} \bspUp{0} \end{tikzpicture} };
	\node (s4) [below = of g4] { \begin{tikzpicture} \bspDown{0} \end{tikzpicture} };
	\node (s6) [below = of g6] { \begin{tikzpicture} \bspUp{0} \end{tikzpicture} };
	\node (s3) [above = of g3, draw=red, circle, yshift=-0.3em] { \begin{tikzpicture}[baseline, yshift=0.0em] \bspNone{0} \end{tikzpicture} };
	\node (s5) [above = of g5] { \begin{tikzpicture} \color{red} \bspUpDown{0} [baseline, yshift=-0.65em] \color{black} \bspNone{0} \end{tikzpicture} };
	\node (s7) [above = of g7] { \begin{tikzpicture} \bspDown{0} \end{tikzpicture} };
	\node (t7) [right = of s6, draw=black] { $E=U-\Delta$ };
\end{scope}
\end{tikzpicture} \\[2em]
}
{
	\includegraphics{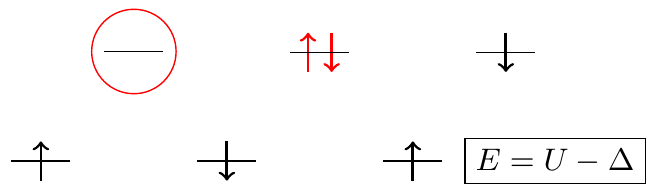}
}
\flushleft{(c) doublon moving to the right}
\ifthenelse{\boolean{buildtikzpics}}
{
	\tikzsetnextfilename{img/tikz_hop_doubr}
	\begin{tikzpicture}
	\begin{scope} [node distance = 0.1 and 0.1]
	\node (g2) { \begin{tikzpicture} \bspGhost{0} \end{tikzpicture} };
	\node (g3) [right = of g2] { \begin{tikzpicture} \bspGhost{0} \end{tikzpicture} };
	\node (g4) [right = of g3] { \begin{tikzpicture} \bspGhost{0} \end{tikzpicture} };
	\node (g5) [right = of g4] { \begin{tikzpicture} \bspGhost{0} \end{tikzpicture} };
	\node (g6) [right = of g5] { \begin{tikzpicture} \bspGhost{0} \end{tikzpicture} };
	\node (g7) [right = of g6] { \begin{tikzpicture} \bspGhost{0} \end{tikzpicture} };
	\node (g8) [right = of g7] { \begin{tikzpicture} \bspGhost{0} \end{tikzpicture} };
	\node (s2) [below = of g2] { \begin{tikzpicture} \bspUp{0} \end{tikzpicture} };
	\node (s4) [below = of g4] { \begin{tikzpicture} \bspDown{0} \end{tikzpicture} };
	\node (s6) [below = of g6] { \begin{tikzpicture} \color{red} \bspUpDown{0} [baseline, yshift=-0.65em] \color{black} \bspNone{0} \end{tikzpicture} };
	\node (s3) [above = of g3, draw=red, circle, yshift=-0.3em] { \begin{tikzpicture}[baseline, yshift=0.0em] \bspNone{0} \end{tikzpicture} };
	\node (s5) [above = of g5] { \begin{tikzpicture} \bspUp{0} \end{tikzpicture} };
	\node (s7) [above = of g7] { \begin{tikzpicture} \bspDown{0} \end{tikzpicture} };
	\node (t7) [right = of s6, draw=black] { $E=U$ };
\end{scope}
\end{tikzpicture} \\[2em]
}
{
	\includegraphics{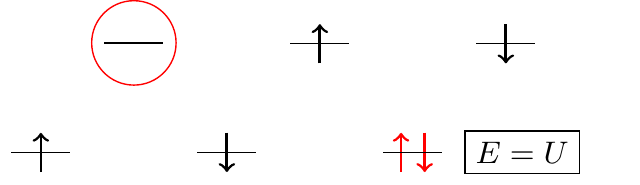}
}
\flushleft{(d) doublon moving to the left}
\ifthenelse{\boolean{buildtikzpics}}
{
	\tikzsetnextfilename{img/tikz_hop_doubl}
	\begin{tikzpicture}
	\begin{scope} [node distance = 0.1 and 0.1]
	\node (g2) { \begin{tikzpicture} \bspGhost{0} \end{tikzpicture} };
	\node (g3) [right = of g2] { \begin{tikzpicture} \bspGhost{0} \end{tikzpicture} };
	\node (g4) [right = of g3] { \begin{tikzpicture} \bspGhost{0} \end{tikzpicture} };
	\node (g5) [right = of g4] { \begin{tikzpicture} \bspGhost{0} \end{tikzpicture} };
	\node (g6) [right = of g5] { \begin{tikzpicture} \bspGhost{0} \end{tikzpicture} };
	\node (g7) [right = of g6] { \begin{tikzpicture} \bspGhost{0} \end{tikzpicture} };
	\node (g8) [right = of g7] { \begin{tikzpicture} \bspGhost{0} \end{tikzpicture} };
	\node (s2) [below = of g2] { \begin{tikzpicture} \bspUp{0} \end{tikzpicture} };
	\node (s4) [below = of g4] { \begin{tikzpicture} \color{red} \bspUpDown{0} [baseline, yshift=-0.65em] \color{black} \bspNone{0} \end{tikzpicture} };
	\node (s6) [below = of g6] { \begin{tikzpicture} \bspUp{0} \end{tikzpicture} };
	\node (s3) [above = of g3, draw=red, circle, yshift=-0.3em] { \begin{tikzpicture}[baseline, yshift=0.0em] \bspNone{0} \end{tikzpicture} };
	\node (s5) [above = of g5] { \begin{tikzpicture} \bspDown{0} \end{tikzpicture} };
	\node (s7) [above = of g7] { \begin{tikzpicture} \bspDown{0} \end{tikzpicture} };
	\node (t7) [right = of s6, draw=black] { $E=U-2\Delta$ };
\end{scope}
\end{tikzpicture} \\[2em]
}
{
	\includegraphics{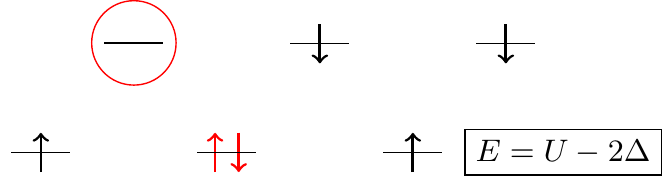}
}
\end{center}
\caption
{
	Illustration of the effect of a spin-selective excitation \eqref{eq:lambda}, in which only the spin-$\uparrow$ electrons are affected. 
	We depict only one contribution which moves the electron by one lattice site. 
	(a) Situation immediately after the excitation. 
	The numbers at each position give the energy cost in the atomic limit $\hopping = 0$, which sum up to the total energy $E$. 
	(b) The hole has hopped to the right resulting in an increased total energy.
	(c) Situation in case the doublon moves away from the hole: The total energy increases further. 
	(d) Situation in case the doublon moves towards the hole: $E$ decreases again. 
} 
\label{fig:sketch_excitations}
\end{figure}
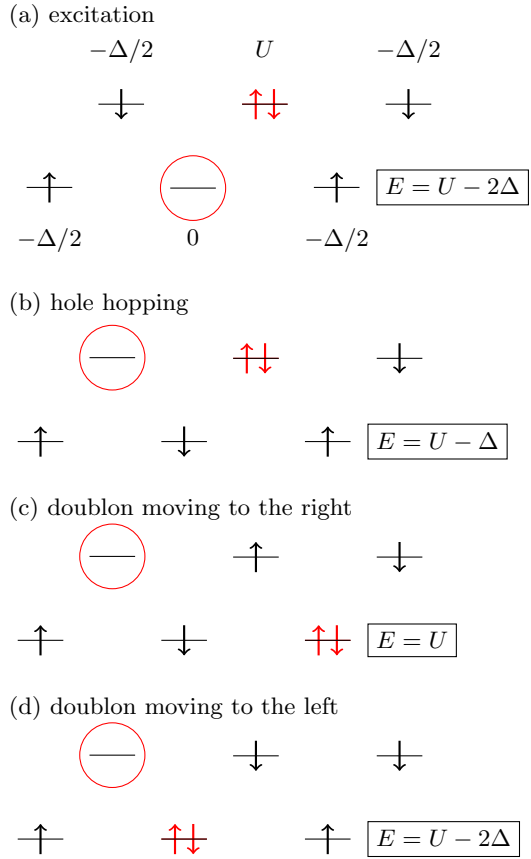
\subsubsection{Exciton confinement}
The findings are explained by the alternating magnetic background potential, which hinders the motion of the hole and of the doublon after the excitation, as illustrated in Fig.~\ref{fig:sketch_excitations} for a single localized excitation for $\hat{H}^{2 \Delta}$:
At finite $U/\hopping$, the $\downarrow$-electrons after the excitation are on the one hand repelled by the excited $\uparrow$-electron and would favor to move away, on the other hand they are hindered by the staggered potential in their motion, so that an effective binding to the original place is realized. 
Furthermore, at half filling the energy of the system grows with the separation $d$ of the hole and the doublon as $d \cdot \Delta$. 
This leads to a confinement of doublon and hole, since the energy is lowest if both are neighboring each other, and hence to the formation of an exciton. 
Note that the exciton is able to move through the system, since the energy remains the same as long as doublon and hole sit next to each other. 
The $\uparrow$-electron can furthermore not directly move back to the original place, since it has no channel to distribute the energy gained after the excitation, see the detailed discussion in Ref.~\onlinecite{koehler2020formation_published}, where this mechanism lead to the formation of long-lived spatial density patterns.
Such a recombination process is only possible when scattering to further particles takes place and is studied further in the next section. 

In Ref.~\onlinecite{Bittner2020}, midgap states in the spectral function of correlated insulators were associated to excitons in two-dimensional extended Hubbard systems with nearest neighbor (n.n.) interaction $V$, which is the cause for the binding between hole and doublon.
In comparison, the exciton observed here has some unusual properties: i) in contrast to Refs.~\onlinecite{Excitons2001,Jeckelmann2003} the excitonic signature in $\sigma(\omega,t)$ appears only \textit{after} the photoexcitation, indicating this is a dark exciton\cite{Selig_excitons,Robert_dark_excitons,jiang_real_time_exciton}; ii) the exciton is formed without a $V$-term in the Hubbard Hamiltonian, i.e., even at very strong screening; iii) the features differ for both spin directions; iv) it appears at a different energy. 

The study of exciton signatures in ARPES is an ongoing topic \cite{perfetto_melting_noneq_exc,wallauer_momentum_observation,dong_measurement_exciton,stefanucci_arpes_exciton,madeo_visualizing_dark_exc}. 
In these experiments, an electron is emitted, and hence only the breaking of an exciton can be observed, i.e. one needs to photoexcite the system first (as in pump-probe setups) to create the exciton.
An in-gap feature below the conduction or upper Hubbard band, respectively for band or Mott insulators, is expected to appear, which is at a position lower by $E_b$ than the edge of the upper band~\cite{christiansen2019,perfetto_melting_noneq_exc}. 
Here, however, the exciton feature in the spectral function appears at an energy $E_b$ {\it above} the lower Hubbard band.
This is the case, since the upper Hubbard band in the $\downarrow$-direction remains empty even after the excitation. 
\begin{figure}
	\centering
	\includegraphics[width=0.48\textwidth]{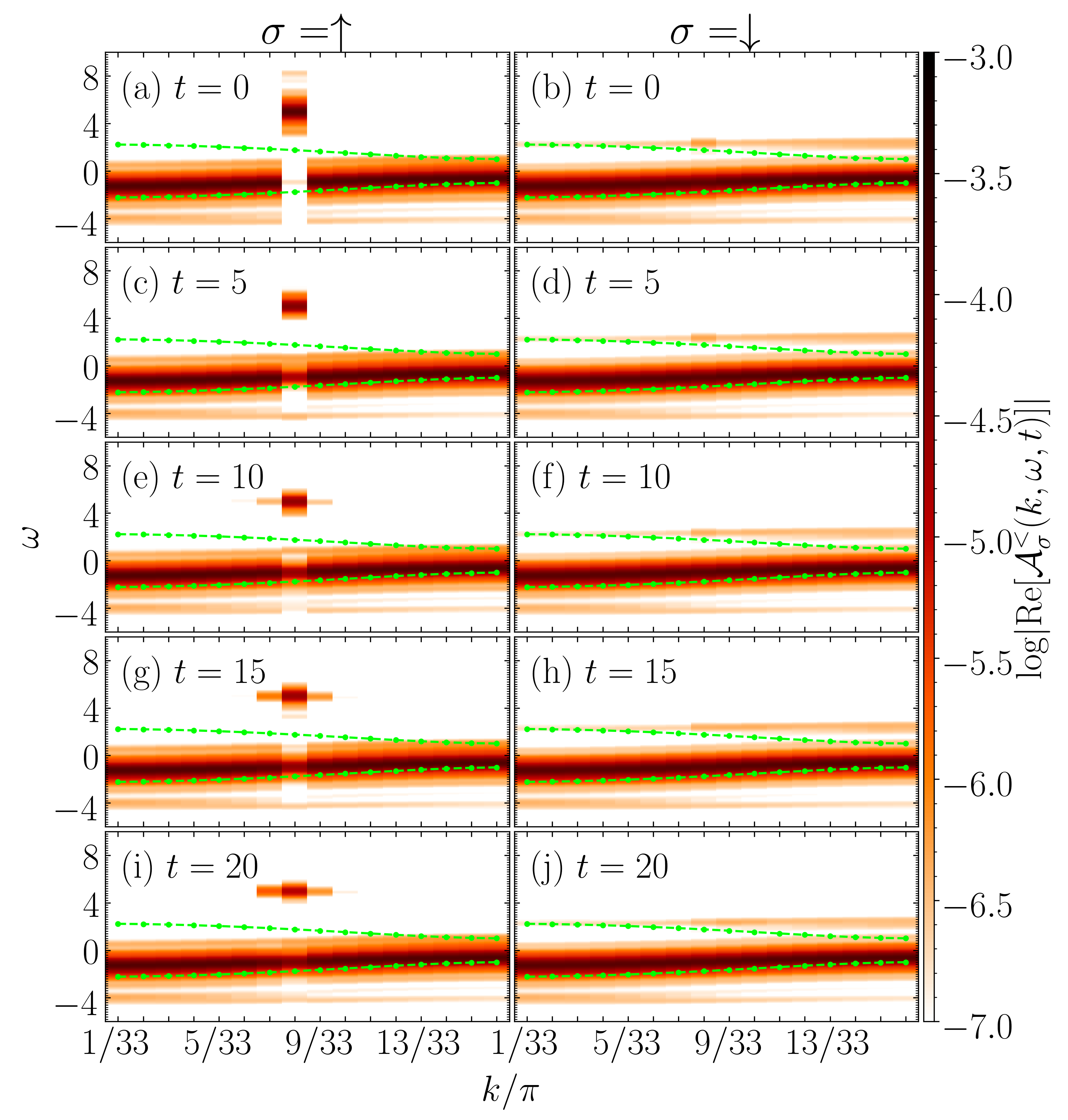}
	\caption{$\mathcal{A}^{<}_{\sigma}\!\left(k,\omega,t\right)$ at different times, $t = 0$ being directly after the electron-hole excitation in the $\uparrow$-direction at $k = 8\pi/33$, $U/t_{\mathrm{h}} = 4$,  $L = 32, \Delta/t_{\mathrm{h}} = 2$ for $\hat{H}^{2\Delta}$ at half filling in the first BZ, c.f. Figs.~\ref{fig:m_ex_t_0_u_cen_2pcmo}. 
		Left (right) column: result for the $\uparrow$-electrons ($\downarrow$-electrons). 
		Note that in the $\uparrow$-direction the position of the excited particle corresponds to populating the lower branch of $\mathcal{A}^{>}_{\uparrow}\!\left(k,\omega\right)$ at the value of $k$ of interest. 
		In contrast, in the $\downarrow$-direction a population inside the gap is obtained, which is smeared out over the entire reduced BZ.
		The green dashed lines show the band structure of the non-interacting system 
		calculated with PBC, the green dots correspond to the calculation with OBC.
	}
	\label{fig:movie_cen_2pcmo}
\end{figure}
\begin{figure}
	\includegraphics[width=0.48\textwidth]{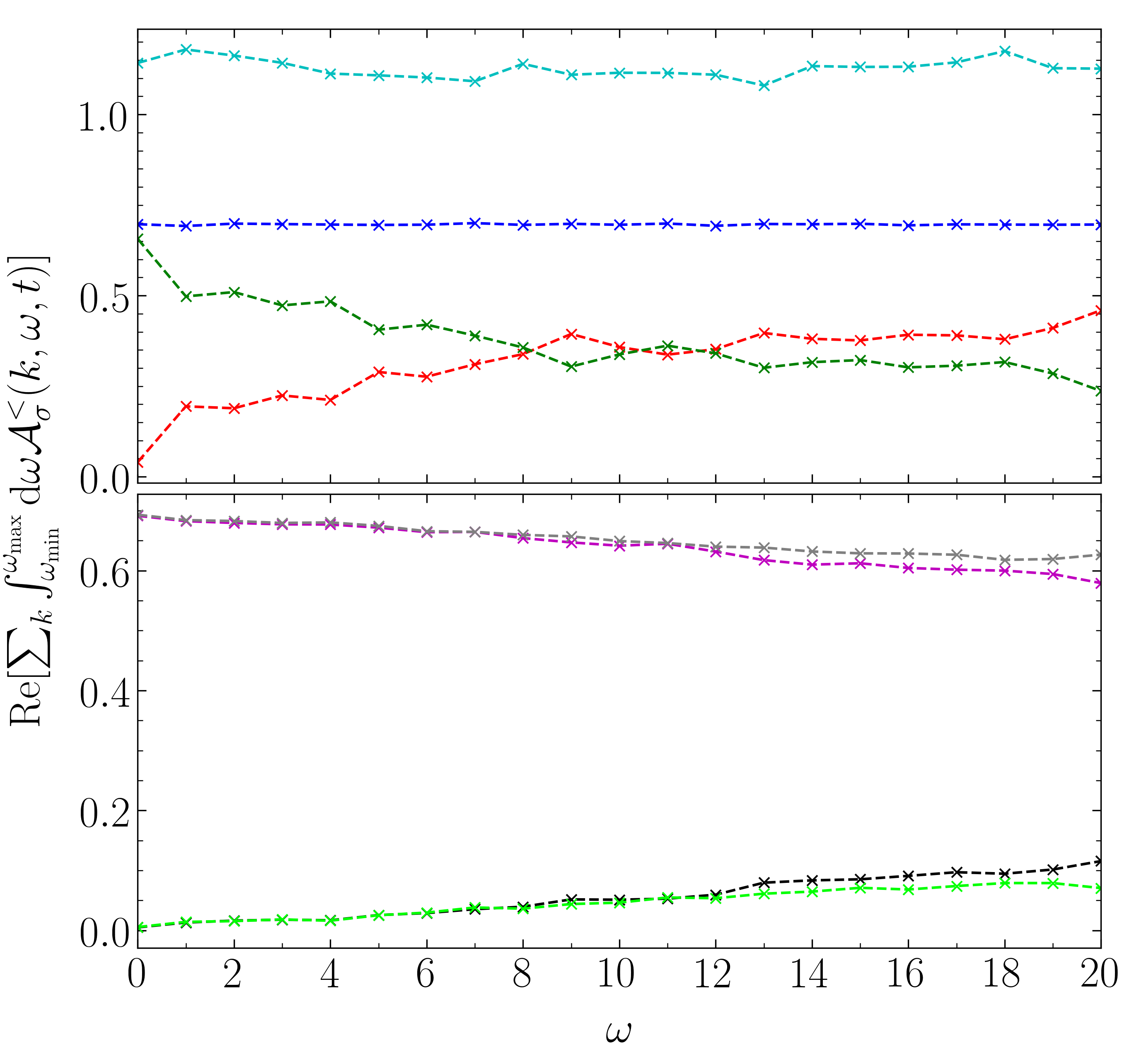}
	\caption{Total spectral weight of $\mathcal{A}^{<}_{\sigma}\!\left(k,\omega,t\right)$ in between frequencies $\omega_{\mathrm{min}}$ and $\omega_{\mathrm{max}}$ summed over certain momenta $k$ for $\hat{H}^{2\Delta}$ at half filling after the electron-hole excitation as specified in Fig.~\ref{fig:m_ex_t_0_u_cen_2pcmo}. 
		Top panel: $\sigma = \uparrow$, $\omega_{\mathrm{min}} \approx -2.0$, $\omega_{\mathrm{max}} \approx 0.0$, $k = 8\pi/33$: red; $\sigma = \uparrow$, $\omega_{\mathrm{min}} \approx 4.0$, $\omega_{\mathrm{max}} \approx 6.0$, $k = 8\pi/33$: green; sum of red and green line: blue; $\sigma = \downarrow$, $\omega_{\mathrm{min}} \approx 1.3$, $\omega_{\mathrm{max}} \approx 4.0$: cyan. Bottom panel: $\sigma = \uparrow, \omega_{\mathrm{min}} \approx -2.0, \omega_{\mathrm{max}} \approx 0.0, k=7\mathrm{\pi}/33$: magenta; $\sigma = \uparrow, \omega_{\mathrm{min}} \approx 4.0, \omega_{\mathrm{max}} \approx 6.0, k=7\mathrm{\pi}/33$: black; $\sigma = \uparrow, \omega_{\mathrm{min}} \approx -2.0, \omega_{\mathrm{max}} \approx 0.0, k=9\mathrm{\pi}/33$: gray; $\sigma = \uparrow, \omega_{\mathrm{min}} \approx 4.0, \omega_{\mathrm{max}} \approx 6.0, k=9\mathrm{\pi}/33$: light green. 
	}
	\label{fig:weights_2pcmo_closeup}
\end{figure}
Hence, removing a $\downarrow$-electron in the photoemission process will cost the energy given by the band structure plus the binding energy of the exciton, leading to the feature at the observed energy.
The same would also be true if the excitation was not only acting on one spin-direction, so that the spectral function will show such a feature at this energy also in this case.
This has an interesting consequence: in correlated band insulators as the ones treated here, trARPES measurements can obtain an additional feature not at the energy expected for bright excitons, but at the energy given by the lower Hubbard band plus the binding energy.

The discussion in this section leads us to conclude that an on-site Hubbard interaction $U$ in the presence of a magnetic superstructure generically can lead to the formation of dark excitons in the spin direction opposite to the excited one, which is further supported by the similarity of our findings for the two different models $\hat{H}^{4\Delta}$ and $\hat{H}^{2 \Delta}$. 

\subsection{Transient behavior}
\label{sec:transient}

Now we analyze the time evolution of the excited system.
We focus on the behavior of $\hat{H}^{2\Delta}$, whose time evolution shows essentially the same behavior as the one of $\hat{H}^{4 \Delta}$ but is easier to discuss.
In Fig.~\ref{fig:movie_cen_2pcmo} we display $\mathcal{A}^{<}_{\sigma}\!\left(k,\omega,t\right)$ at selected times $t$, including $t=0$ directly after the excitation corresponding to Figs.~\ref{fig:m_ex_t_0_u_cen_2pcmo}(a) and (b).
Comparing Figs.~\ref{fig:movie_cen_2pcmo}(a) and (i), in the $\uparrow$-direction we find the main effect is a recombination of the electron-hole pair.
However, also at neighboring $k$-values small effects are visible.
To further analyze this, we display in Fig.~\ref{fig:weights_2pcmo_closeup} the time evolution of the populations in various $(k,\omega)$-regions of interest (ROI) in the spectral function. 
The red and the green lines show the population in the ROI to which the particle was excited to or from which it was taken from ($k = 8 \pi/33$ and approximately $\omega \in [-2,0]$ or $\omega \in [4,6]$, respectively). 
The blue line shows the sum of the both.
We find that the populations in the ROIs are $\sim 0.1$ or $\sim 0.7$, respectively, at time $t=0$.
As discussed previously, for a non-interacting system we would expect populations of $0$ and $1$, and the deviation is due to correlation effects.
The interactions induce a recombination of this electron-hole pair, and we see that on a time scale $\sim 10$ the weight of both regions becomes the same.
The recombination process continuous, but will take much longer than the time scales treated by us.
The sum of both weights is approximately constant in time, but shows small fluctuations.
Note that its value is different from 1, since there are small weights at the same $k$-value also outside these ROIs (taking these into account we find the contributions indeed sum up to 1). 
Since the sum of both contributions is approximately constant, scattering from or to other  $k$-values seems to play a minor role. 
Indeed, we find that for all $k$-values further away the populations within our estimated accuracy do not change in time.
However, for the neighboring $k$-values, we find that interband scattering leads to a redistribution of weights from the lower to the upper Hubbard band in the course of time, which leads to a change linear in time of the populations in these ROIs on the time scales investigated by us, see Fig.~\ref{fig:weights_2pcmo_closeup}.

\begin{figure*}
	\includegraphics[width=\textwidth]{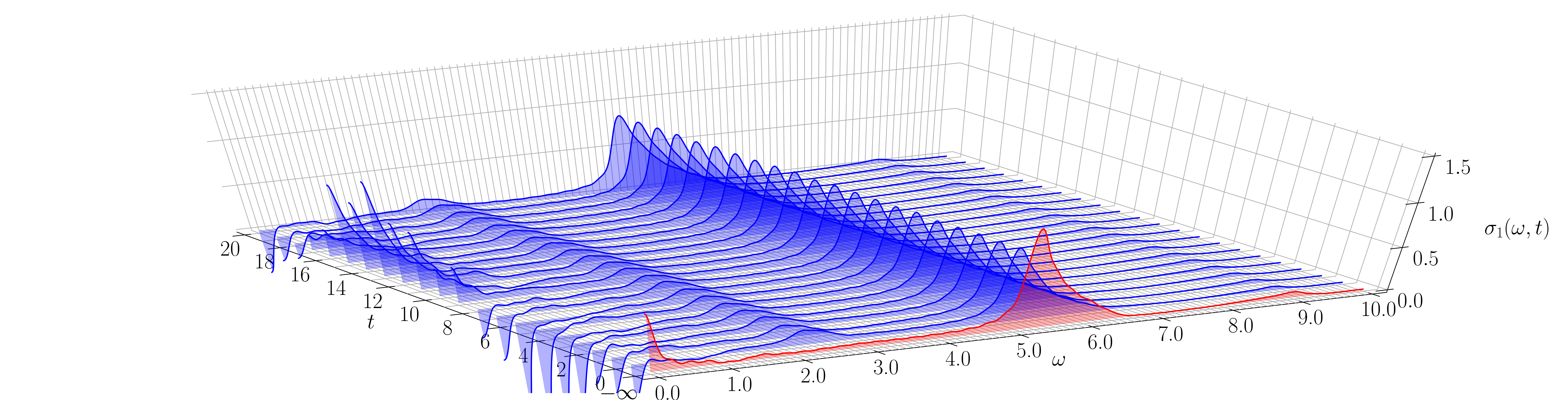}
	\caption{Real part of the time-dependent optical conductivity $\sigma_{1}\!\left(\omega,t\right)$ for $\hat{H}^{2\Delta}$ with $\Delta/\hopping=2$ and $U/\hopping=4$ at half filling before (ground state, $t=-\infty$, red; c.f. Fig.~\ref{fig:oc_gs_vs_t_0_2}(b)) and after ($t \geq 0$, blue) the excitation \eqref{eq:lambda}. The data was damped using $\tilde{\eta}=0.1$ and $16$ times zero padding was applied in \eqref{eq:methods_oc_ft_a} and \eqref{eq:methods_oc_ft_j}.
	}
	\label{fig:oc_plot_u_4_d_2_h_gs_vs_t_2pcmo}
\end{figure*}
In the $\downarrow$-direction, the populations do essentially not change in time on the time scales treated by us.
Interestingly, this is also true for the total weight of the excitonic band, which is approximately $1$. 
This indicates that the excitonic state has a lifetime substantially longer than the time scale investigated. 

We complement this discussion by the time-dependence of the optical conductivity for $\hat{H}^{2\Delta}$ shown in Fig.~\ref{fig:oc_plot_u_4_d_2_h_gs_vs_t_2pcmo}. 
We focus in particular on the features induced by the excitation finding that the additional peak at $\omega \approx 2$ is stable and does essentially not change with time.
We see that the other features at frequencies $\omega \gtrsim 1$ also do not change in time, and that no further features appear.
At low frequencies, we observe a time-dependent oscillation at $\omega \to 0$; however, as discussed in the method section, we believe that the approach is not accurate enough to make precise statements about this behavior and leave this to future research. 

\section{Conclusion and Outlook}
\label{sec:outlook}

We investigated the time evolution of one-dimensional Hubbard-like systems at half filling with a magnetic superstructure following a spin-selective electron-hole excitation studying the spectral function and the optical conductivity in and out-of-equilibrium. 

In a first step, we treat the ground state spectral function and identify at finite interactions $U$ an upper and a lower Hubbard band, which posses an additional fine structure caused by the super structure.
An FD calculation for a system consisting of only a single unit cell confirmed our MPS-obtained results. 
In the ground state optical conductivity we find peaks which can be identified with band transitions in the spectral function. 

Afterwards, we computed the time evolution of two variants of the system after an excitation in the spin-$\uparrow$ direction only. 
We observed recombination of the excited electron and the hole in $\mathcal{A}_{\uparrow}^{<}\!\left(k,\omega,t\right)$.
At $t \sim 10$ we found the populations to have become equally strong. 
In addition, at neighboring $k$-values inter band scattering leads to a roughly linear population growth in the upper Hubbard band in the spin-$\uparrow$ direction. 
For $\hat{H}^{4\Delta}$ we do find an additional peak in $\sigma_{\uparrow,1}\!\left(\omega,t\right)$ in the gap region. 
However, this feature is also found for $U=0$ and, hence, does not indicate the formation of an exciton but is due to the complex band structure. 
Indeed, for the simpler band structure of $\hat{H}^{2\Delta}$ this feature disappears. 
Thus, in the spin-$\uparrow$ direction no indication of exciton formation is obtained. 

However, we found for $U>0$ in all cases an in-gap signal in the spin-$\downarrow$ direction even though it was not touched by our excitation. 
In the optical conductivity additional peaks are realized only in the $\downarrow$-direction, which we associate with the features in the spectral function.
Since this is obtained only for $U>0$, we can rule out the super structure as the sole cause of this effect. 
Furthermore, for a doublon-hole pair the energy increases linearly with separation, i.e. there is a confinement of the doublon-hole pair to nearest neighbors. 
At $U>0$ this leads to a finite binding energy, i.e. excitons are formed, which appear only after the photo excitation and only in the opposite spin direction, which we therefore call spinful dark excitons.
In the spectral function these excitons form a band, which is at the value of the binding energy above the edge of the lower Hubbard band. 
Note that in Mott insulators multiple excitons can be formed\cite{Excitons2001} which we see in Figs.~\ref{fig:a_oc_gs_vs_t_0_4_delta_8} and \ref{fig:oc_gs_vs_t_0_4_delta_8}.

In the time evolution we find these new features to be nearly independent of time in both the spectral function and the optical conductivity on the time scales treated by us. 

It would be interesting to investigate for such effects in materials where magnetic super structures are realized.
Examples are CE-structures in manganites\cite{Hotta2004}, or orbital-selective Mott phases in iron-based ladder compounds such as $\text{BaFe}_{2}\text{Se}_{3}$\cite{Jacek_2020,Jacek_2020_2,Jacek_2019,Jacek_2021}. 
Alternatively, this can also be studied in ultracold gases on optical lattices\cite{Bloch:2005p988,Bloch:2008p943,Bloch2012}, on which it is possible to realize superlattices\cite{superlattice_magneticfield,superlattice_magneticfield_PRL} and to investigate for spectral functions.\cite{Jin2008}

\begin{acknowledgements}
The authors thank B. Fauseweh, A. Osterkorn, K. Harms, F. Sohn, D. Jansen, J. Stolpp, M. Hopjan, S. Paeckel, T. K\"{o}hler, M. Kalthoff, P. Bl\"{o}chl, F. Heidrich-Meisner, F. Gebhard, R.M. Noack, and E. Arrigoni for fruitful discussions. 
We are grateful for many stimulating and insightful discussions with all participants of the journal club of the B07 project of the SFB 1073, in particular also S. Mathias and M. Reutzel. 
The work was supported by the North-German Supercomputing Alliance (HLRN). 
We are grateful to the HLRN supercomputer staff, especially S. Krey.
We also acknowledge access to computational resources provided by the GWDG, as well as technical assistance by S. Krey and M. Boden. 
This work is funded by the Deutsche Forschungsgemeinschaft (DFG, German Research Foundation) - 217133147/SFB 1073, project B03.
The results presented in this work were generated using the SymMPS toolkit\cite{symmps}.
\end{acknowledgements}
\appendix 
\section{Equilibrium spectral functions at larger $U$}
\label{app:u}
 
In order to further investigate the impact of $U$ on the band structure, we give a more detailed analysis at this point. 
We begin with $\hat{H}^{4 \Delta}$, where in Sec.~\ref{sec:u} we have already analyzed the effect $U$ has on $\mathcal{A}^{<}_{\uparrow}\!\left(k,\omega\right)$ at half filling, c.f. Fig.~\ref{fig:u_lehmann_l}. 
\begin{figure}[b!]
	\centering
	\includegraphics[width=0.48\textwidth]{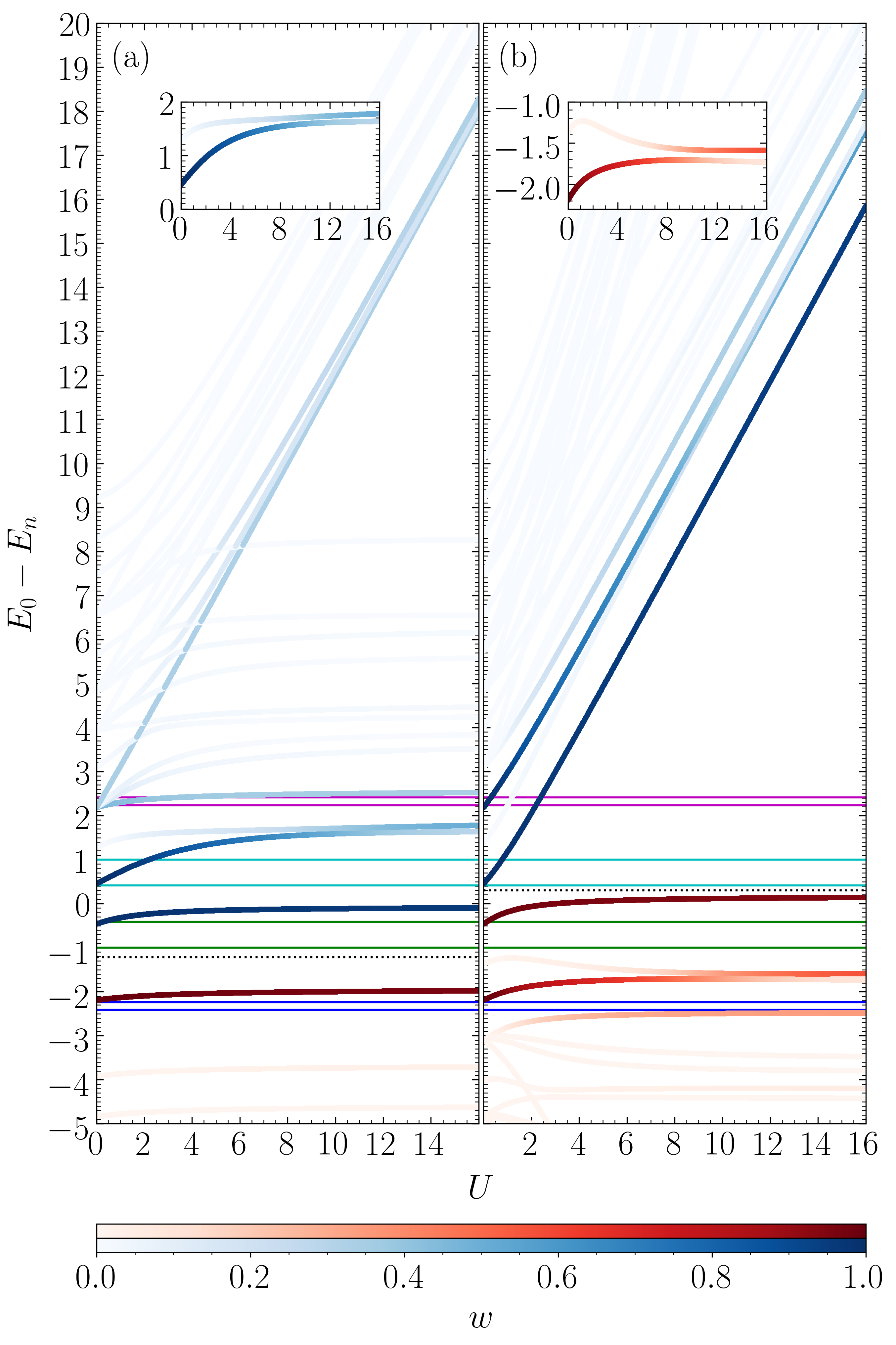}
	\caption{
	$\mathcal{A}^{<}_{\uparrow}\!\left(k,\omega\right)$ (red tones) and $\mathcal{A}^{>}_{\uparrow}\!\left(k,\omega\right)$ (blue tones) as a function of $U$ obtained from FD of a single unit cell of $\hat{H}^{4 \Delta}$ with OBC according to Eq.~\eqref{eq:lehmann_calc} for the case of quarter (a) and half filling (b). The intensity of the line color indicates the weight $w$ of the respective peaks $w_{n}\!\left(U\right) = \sum_{r} \left|\left<n\right|\hat{c}_{\uparrow,r}\left|\mathrm{GS}\right>\!\left(U\right)\right|^{2}$. The figure was cut at $E_{0}-E_{n}=-5$ at the bottom and $E_{0}-E_{n}=20$ at the top when all peaks occur with $w \ll 0.2$. Insets: Zoom into the specified regions. Dashed lines: Separator of $\mathcal{A}^{<}_{\uparrow}\!\left(k,\omega\right)$ and $\mathcal{A}^{>}_{\uparrow}\!\left(k,\omega\right)$. Solid lines: Analytical results for the edges of bands $\nu=1$ (blue), $\nu=2$ (green), $\nu=3$ (cyan), and $\nu=4$ (magenta) for $U=0$ and PBC. Note that some lines with finite weights at $U=0$ are not located in the respective bands which is due to their computation with OBC. Fig.~\ref{fig:u_lehmann}(b) also shows data from Fig.~\ref{fig:u_lehmann_l}.
	}
	\label{fig:u_lehmann}
\end{figure}
Fig.~\ref{fig:u_lehmann}(b) additionally depicts $\mathcal{A}^{>}_{\uparrow}\!\left(k,\omega\right)$, illustrating the symmetry of $\mathcal{A}^{<}_{\uparrow}\!\left(k,\omega\right)$ and $\mathcal{A}^{>}_{\uparrow}\!\left(k,\omega\right)$ at $U/2$ as expected from particle-hole symmetry. 
Clearly, the weight takeover, which is highlighted in the inset, also occurs for $\mathcal{A}^{>}_{\uparrow}\!\left(k,\omega\right)$, which is best seen at $U/\hopping=14$. 
From the inset, we find that at $U/\hopping \approx 8$ the weights of the overtaking and overtaken line are roughly equal, while at $U/\hopping=16$ the process is more or less completed. 

Interestingly, we find a similar behavior also at quarter filling as is shown in Fig.~\ref{fig:u_lehmann}(a). 
Here, there is only little change to $\mathcal{A}^{<}_{\uparrow}\!\left(k,\omega\right)$ with growing $U$ as expected due to the low particle density. 
While it comes only with a very small renormalization and three almost negligible side bands, $\mathcal{A}^{>}_{\uparrow}\!\left(k,\omega\right)$ exploits a much richer structure. 
In addition to a similar weight takeover to the third band as in the case of half filling, which is highlighted in the inset, the fourth band showcases another interesting phenomenon: 
At $U=0$ one expects to find a sole signal of weight $1$ analytically, but when looking into the numbers, we find in fact two signals of weight $0.5$ each. 
Hence, two signals appear to have collapsed into one resulting in a degeneracy. 
This shows an actual band splitting when increasing $U/\hopping$, unlike the weight take over we discussed above.
In the latter case, it is not clear from our computations that it must be a contribution from the same band the weight from a fading-out line is transferred to. 
The degeneracy, however, leaves no other explanation than an actual splitting.
Note that only one of the two major branches from the split fourth band scales with $U$, the other one is only slightly renormalized around $\omega \approx 2.5$.
Again, the former branch is subject to weight takeover for larger $U/\hopping$.  

For $\hat{H}^{2 \Delta}$, the spectral function for a single unit cell gained from \eqref{eq:lehmann_calc} is rather simple consisting of only two contributions, a dominant band, which survives in the case $U=0$, and a sub band, which takes only some weight from the dominant one for large $U$, c.f. Fig.~\ref{fig:u_lehmann_l_2pcmo}.
Both contributions are also subject to some renormalization and $\mathcal{A}^{>}_{\uparrow}\!\left(k,\omega\right)$ will scale with $U$, making a detailed analysis as in Fig.~\ref{fig:u_lehmann} obsolete. 
The spectral functions simply flattens out considerably for large $U$ such that the comparison to the analytical computation using the Lehmann representation will agree neatly to the MPS data.
As for the case of $\hat{H}^{4 \Delta}$ we may interpret the system as being composed of two Hubbard bands with a fine structure, which due to the considerably simpler structure of $\hat{H}^{2 \Delta}$ is less evolved consisting of only a sub band for both $\mathcal{A}^{<}_{\uparrow}\!\left(k,\omega\right)$ and $\mathcal{A}^{>}_{\uparrow}\!\left(k,\omega\right)$. 

\section{Transformation to quasi-momenta in case of OBC}
\label{app:obc}

Concerning the Fourier transform to $k$-space of the spectral functions as presented in Sec.~\ref{sec:methods_spectralfcts}, it is most straight forward to implement it using periodic boundary conditions (PBC).
In our case with an extended unit cell this results to a transform in a modified plain-wave basis, which are the eigenstates of the non-interacting Hamiltonian in one unit cell.
However, MPS work best for systems with open boundary conditions (OBC), for which some care needs to be taken.
As pointed out in Refs.~\onlinecite{delplace_zak_phase, Marques_2020,Matulis2009}, this can be rather involved and non-trivial effects, e.g. Zak phases, can come into play.
While these are interesting aspects, they lie outside the focus of our study, so that we want to identify the most direct way to obtain a transformation to (quasi-)momenta in the case of OBC, such that in the non-interacting case the time-dependent spectral function does not change in time. 

For simple systems, a transform using $\sin$-functions is useful\cite{benthien_sin}; however, for systems with an extended unit cell this can be more involved and it can be very cumbersome to write down the generalized plain-wave basis analytically.
Therefore, instead, we numerically diagonalize the non-interacting Hamiltonian of the entire system and transform into its eigenbasis.
As described in more detail in Ref.~\onlinecite{koehler2020formation_published}, this results in the generalized Fourier transform 
\begin{equation}
	\mathcal{G}^{<}_{\sigma}\!\left(k,t^{\prime},t\right) = \sum_{r^{\vphantom{\prime}},r^{\prime}}P^{\sigma}_{r^{\vphantom{\prime}},k}P^{\sigma *}_{r^{\prime}\!,k}\mathcal{G}^{<}_{\sigma}\!\left(r^{\vphantom{\prime}},r^{\prime},t^{\prime},t\right)\!.
\end{equation}
The matrices $P^{\sigma}_{r,k}$ are obtained from the diagonalization (for $U=0$) of \eqref{eq:model_pcmo} with OBC, through the definition of the annihilation and creation operators 
\begin{equation}
	\hat{a}^{\vphantom{\dagger}}_{\sigma,k^{\vphantom{\prime}}} = \sum_{j}P_{j,k^{\vphantom{\prime}}}^{\sigma *}\hat{c}^{\vphantom{\dagger}}_{\sigma, j\vphantom{+1}}, \quad \text{and} \quad \hat{a}^{\dagger}_{\sigma,k^{\vphantom{\prime}}} = \sum_{j}P_{j,k^{\vphantom{\prime}}}^{\sigma}\hat{c}^{\dagger}_{\sigma, j\vphantom{+1}}.
\end{equation}
Here, we choose the hermitian matrices $P^{\sigma}$ such that they hold the eigenvectors of the non-interacting Hamilton matrix $H^{\sigma}$ given by
\begin{equation}
	\hat{H}_{0} = \sum_{\sigma} \sum_{i,j} H_{i,j}^{\sigma} \hat{c}^{\dagger}_{\sigma, i\vphantom{+1}}\hat{c}^{\vphantom{\dagger}}_{\sigma, j\vphantom{+1}},
\end{equation}
and, thus, 
\begin{equation}
\label{eq:app_obc_gen_sin}
	H^{\sigma} = {P^{\sigma}} D^{\sigma} P^{\sigma\dagger} \Leftrightarrow H^{\sigma}_{i,j} = \sum_{m} P^{\sigma}_{i,m} D^{\sigma}_{m,m} P^{\sigma *}_{j,m}.
\end{equation} 
The matrices $D^{\sigma}$ are diagonal holding the eigenvalues of $H^{\sigma}$. 
Renaming $m \rightarrow k$ and $D_{m,m}^{\sigma} \rightarrow \epsilon^{\sigma}\!\left(k\right)$ we may identify them with the system's dispersion relation. 
Note that similar to the sine transform, this method only gives half of the Brillouin zone, i.e.
\begin{equation}
	\label{eq:model_momenta}
	k = \frac{\pi p}{L+1}, \quad p \in \left\{1, \dots , L\right\}.
\end{equation}
For $L \rightarrow \infty$ and $k > 0$ the newly defined $\epsilon^{\sigma}\!\left(k\right)$ will converge to Eq. \eqref{eq:model_disp_4} or \eqref{eq:model_disp_2}, respectively. 
We further stress that the procedure outlined above produces results in the extended zone scheme. 
One may, however, fold back manually to the first Brillouin zone defining corresponding momentum space operators for OBC through
\begin{equation}
	\label{eq:model_fold_back_4}
	\hat{a}^{\left(\dagger\right)}_{\sigma,\nu,k^{\prime}} =
	\begin{cases}
		\hat{a}^{\left(\dagger\right)}_{\sigma,k^{\prime}},& \nu = 1 \\
		\hat{a}^{\left(\dagger\right)}_{\sigma,\left(L/2+1\right)\pi/\left(L+1\right)-k^{\prime}} ,& \nu = 2 \\
		\hat{a}^{\left(\dagger\right)}_{\sigma,\left(L/2\right)\pi/\left(L+1\right)+k^{\prime}} ,& \nu = 3 \\
		\hat{a}^{\left(\dagger\right)}_{\sigma,\pi-k^{\prime}} ,& \nu = 4 \, ,\\
	\end{cases}
\end{equation}
introducing again the band index $\nu$ and calculating the momenta $k^{\prime}$ using \eqref{eq:model_momenta} with $p^{\prime} \in \left\{1, \dots , L/4\right\}$ for $\hat{H}^{4\Delta}_{0}$ and 
\begin{equation}
	\label{eq:model_fold_back_2}
	\hat{a}^{\left(\dagger\right)}_{\sigma,\nu,k^{\prime}} =
	\begin{cases}
		\hat{a}^{\left(\dagger\right)}_{\sigma,k^{\prime}},& \nu = 1 \\
		\hat{a}^{\left(\dagger\right)}_{\sigma,\pi-k^{\prime}} ,& \nu = 2 \, ,\\
	\end{cases}
\end{equation}
for $p^{\prime} \in \left\{1, \dots , L/2\right\}$ in the case of $\hat{H}^{2\Delta}_{0}$.

\bibliographystyle{prsty}
\bibliography{lit}

\end{document}